\documentclass[twocolumn,tighten,twocolappendix]{aastex63}

\usepackage{epsfig}
\usepackage{epstopdf}
\usepackage{color}
\usepackage{natbib}
\usepackage{amsmath}	% Advanced maths commands
\usepackage{amssymb}	% Extra maths symbols

\newcommand{\be}{\begin{equation}}
\newcommand{\ee}{\end{equation}}
	
\newcommand{\cm}{\mathrm{cm}}	
\newcommand{\s}{\mathrm{s}}

\defcitealias{bai_19}{Paper I}

\graphicspath{{./}{figures/}}

\shorttitle{Cosmic ray streaming with ion-neutral damping}
\shortauthors{I. Plotnikov, E. Ostriker \& X.-N. Bai}

\begin{document}
	\title{\Large{Influence of Ion-Neutral Damping on the Cosmic Ray Streaming Instability: MHD-PIC Simulations}}
	
	\correspondingauthor{Illya Plotnikov, Eve C. Ostriker,  Xue-Ning Bai}

	\author[0000-0002-0074-4048]{Illya Plotnikov}
	\affiliation{IRAP, Universit\'e de Toulouse III - Paul Sabatier, OMP, Toulouse, France}
    \email{illya.plotnikov@irap.omp.eu}
	
	\author[0000-0002-0509-9113]{Eve C.~Ostriker}
    \affiliation{Department of Astrophysical Sciences, Princeton University, 4 Ivy Ln.,
	 Princeton, NJ 08544, USA}
    \email{eco@astro.princeton.edu}
	
	\author[0000-0001-6906-9549]{Xue-Ning Bai}
	\affiliation{Institute for Advanced Study, Tsinghua University, Beijing 100084, China}
    \affiliation{Department of Astronomy, Tsinghua University, Beijing 100084, China}
    \email{xbai@mail.tsinghua.edu.cn}

%\maketitle

\begin{abstract}
We explore the physics of the gyro-resonant cosmic ray streaming instability (CRSI) including  the effects of ion-neutral (IN) damping.  This is the main damping mechanism in (partially-ionized) atomic and molecular gas, which are the primary components of the interstellar medium (ISM) by mass.  Limitation of CRSI by IN damping is important in setting the amplitude of Alfv\'en waves that scatter cosmic rays and control galactic-scale transport. Our study employs the MHD-PIC hybrid fluid-kinetic numerical technique to follow linear growth as well as post-linear and saturation phases. During the linear phase of the instability -- where simulations and analytical theory are in good agreement -- IN damping prevents wave growth at small and large wavelengths, with the unstable bandwidth lower for higher ion-neutral collision rate $\nu_{\rm in}$. Purely MHD effects during the post-linear phase extend the wave spectrum towards larger $k$. In  the saturated state, the cosmic ray distribution evolves toward greater isotropy (lower streaming velocity) by scattering off of Alv\'en waves excited by the instability. In the absence of low-$k$ waves, CRs with sufficiently high momentum are not isotropized.  The maximum wave amplitude and rate of isotropization of the distribution function decreases at higher $\nu_{\rm in}$. When the IN damping rate approaches the maximum growth rate of CRSI, wave growth and isotropization is suppressed. Implications of our results for CR transport in partially ionized ISM phases are discussed.
\end{abstract}

\section{Introduction}

 Cosmic Rays (CRs) are a key component of the interstellar medium (ISM). They are potentially important to ISM dynamics including support against gravity, as their energy density is typically comparable to that of the magnetic field as well as the thermal and turbulent energy of the ISM gas \citep[e.g., ][]{spitzer_78, ferriere_01, draine_11,2015ARA&A..53..199G}. Perhaps even  more important dynamically,  given the large scale height of the CR distribution, is the potential for CRs to contribute in driving galactic winds \citep[e.g.][]{ipavich_75,breitschwerdt_91,zirakashvili_96,everett_08,mao_ostriker_18}. CRs are also crucial to ISM microphysics, driving ionization and dissociation as well as providing the primary heating of gas in regions shielded to  UV \citep[e.g.][]{draine_11,2012ApJ...756..157G}. 
 
 The coupling between CRs and the thermal ISM gas occurs through scattering of CRs off  of magnetic waves, preexisting in the turbulent ISM or self-generated by streaming CRs \citep{kulsrud_pearce_69,1969ApJ...156..303W,skilling_71,kulsrud_05,2018AdSpR..62.2731A}. Higher-energy CRs may interact primarily with externally-generated  waves \citep{blasi_2012a}. However, particles of energy $\lesssim$ GeV dominate by number and energy the overall CR  content.
 At the micro-pc scales comparable to the gyro-radius of GeV protons, the  energy  density of the ISM turbulence is too small to provide efficient scattering  (under the assumption of a cascade from directly-observed turbulence at larger scales), and these particles are believed to be mainly scattered by  self-excited waves.  Alfv\'en waves generated by streaming therefore
 bear primary responsibility for scattering the dominant portion of the CR distribution in our own and other galaxies \citep[e.g.][]{2013PhPl...20e5501Z}.

There is an increasing necessity to provide satisfactory microphysical understanding of the CR streaming instability and its  implications for CR transport in different phases of the  ISM. The possible dynamical role of CRs in driving galactic winds has recently led to reconsideration of the traditional assumption that streaming occurs at the Alfv\'en speed in several analytical studies \citep{, wiener_13, recchia_16, zweibel_17}. In parallel, a number of groups have implemented CRs as a fluid in numerical magnetohydrodynamics (MHD) codes \citep[e.g., ][]{yang_12,dubois_16,pakmor_16, pfrommer_17, thomas_pfrommer_18, jiang_oh_18, hopkins_19} in order to study the effect of CRs on galactic wind generation \citep[e.g., ][]{hanasz_13,girichidis_16,girichidis_18, pfrommer_17,  wiener_17b, ruszkowski_17, butsky_18, dashyan_20}, on the global evolution of supernova remnants \citep{pais_18, dubois_19} or on the multiphase ISM structure \citep{bustard_20}. Global simulations testing different models of CR transport coefficients were recently conducted by \citet{hopkins_21a, hopkins_21b}, suggesting that standard models of CR transport are inconsistent with observed constraints  on large  scales.
At the same  time, observations that probe ionization and chemistry in ISM clouds at small scales also place constraints on CR transport \citep[see review of][and references therein]{padovani_20}, e.g. suggesting diffusive rather than free-streaming behavior in the outer layers of molecular clouds \citep{silsbee_19}.  

Fluid implementations of CRs rely on subgrid prescriptions for diffusion and streaming. To date, these subgrid treatments have  relied on empirical estimates of diffusivities, idealized streaming treatments, or simple models based on analytic predictions for growth and dissipation of waves.  Considering the important role of microphysics to CR transport, it is valuable to pursue a deeper investigation that addresses the processes of wave growth, damping, and particle-wave interactions directly. The most important parameter to be obtained from microphysical studies of CR-ISM interactions is the scattering rate of CRs off of Alfv\'en waves, since this enters in determining the diffusivity in traditional treatments \citep[e.g.][]{skilling_71,hanasz_03}, or the rate of change of the CR flux in two-moment methods \citep[e.g.][]{jiang_oh_18,thomas_pfrommer_18}.

The energy-dependent scattering frequency of CRs, $\nu_{\rm s}$, is proportional to  the wave energy density in the  resonant range. The latter quantity depends on the outcome of the cosmic-ray streaming instability  (CRSI) in the local medium. There are different classes of the CRSI, resonant \citep{lerche_67,kulsrud_pearce_69, skilling_71, ginzburg_73, wentzel_74, skilling_75a, berezinskii_90} and non-resonant \citep{bell_04, amato_blasi_09, bykov_13}.  The latter is important only in the near environment around accelerators (such as supernova remnant shocks) where the electrical current of CRs is large, i.e. $(4\pi / c) J_{\rm CR} R_{L,0} / B_0 \gg 1$, where $J_{\rm CR} = e n_{\rm CR} V_D$ is the electrical current of CRs streaming at speed $V_D$  and with typical gyroradius $R_{L,0}$ \citep{amato_blasi_09}. In the general ISM, the  resonant interaction prevails since the density of CRs is very small ($n_{\rm CR} \sim  10^{-10}-10^{-8} {\rm \cm^{-3}}$). Because we are interested in ambient ISM conditions, here we focus exclusively on the resonant instability. 

Wave growth, triggered by the anisotropy of the distribution function (DF) or CR density gradient, is generally in competition with different damping mechanisms depending on the phase of the ISM. In a partially ionized medium the most important wave damping mechanism is from the collisions between ions  and neutrals. The competition of CRSI with ion-neutral (IN) damping was studied in context of molecular clouds \citep[e.g., ][]{kulsrud_cesarsky_71, zweibel_shull_82, everett_zweibel_11, morlino_gabici_15, ivlev_18} or in  the context of particle acceleration at shocks \citep[e.g., ][]{drury_96, bykov_toptygin_05, reville_07, blasi_12, nava_16, brahimi_20}. 

Of course, damping mechanisms other than ion-neutral collisions can be important in hotter and more diffuse phases of the ISM. Other mechanisms that have been  discussed include  non-linear Landau damping  \citep[NLLD][]{lee_volk_73, kulsrud_78} and damping by interactions with turbulence \citep{farmer_goldreich_04, yan_lazarian_11, lazarian_16}. We also note the recently studied damping by charged dust grains that can be important for CRs with energies $<300$GeV \citep{squire_2020}. As discussed in \citet{nava_16, xu_2016, brahimi_20}, in neutral dominated phases of the ISM, where the ionization fraction is small, IN damping is dominant for waves which are resonant with CRs of energies $E<10$ TeV. Here, we consider only these primarily-neutral phases of the ISM; investigation of effects of alternative damping mechanisms is deferred to future work.

In the present work, we investigate the interplay between CRSI and ion-neutral damping during the linear phase of the CRSI, and  also assess the behavior of the system in post-linear and late-time saturated stages of evolution.  We are motivated to obtain better insight into CR  transport in ISM conditions which are largely neutral (this makes up most of the ISM mass), where the ion/neutral fraction is controlled by photo-ionization or ionization by low-energy CRs. 

Our study of CRSI relies on the MHD-PIC numerical method described in \citet{bai_15}  \citep[see also, ][]{reville_bell_12, vanMarle_18, amano_18, mignone_18, lebiga_18}. This method is well adapted to capture the resonant nature of the CR-fluid interaction while allowing for relatively large space and time scale evolution of the system  and the fact that $n_{\rm CR}/n_i$ is extremely small in the parameter range of interest. The gyro-radius of CRs must  be resolved, but smaller scales (such as ion skin depth and electron scales) not need to be resolved. We note several recent studies have alternatively investigated certain aspects of the CRSI with fully-kinetic \citep{holcomb_spitkovsky_18, shalaby_20} or hybrid-kinetic approaches \citep{haggerty_19, schroer_20} that use particle-in-cell (PIC) methods. These approaches have the advantage of resolving small-scale phenomena on electron- (full-PIC) or proton-skin depth (hybrid-PIC) scales, but are in practice limited in the range of $n_{\rm CR}/n_i$ and other parameters that can  be  studied because the thermal ions are treated via PIC. 

In this work, we follow up on \citet{bai_19} (hereafter \citetalias{bai_19}), where the CRSI was studied using the MHD-PIC approach. There, the main findings include: (i) the linear phase of the instability for both resonant and non-resonant branches can be accurately reproduced with our ``$\delta f$''  MHD-PIC method; (ii) the quasi-linear diffusion (QLD) formalism accurately describes temporal changes in the CR DF, except near pitch angle of 90 degrees (the ``$\mu=0$ crossing'' problem); (iii) crossing of $\mu =0$ is mainly due to nonlinear wave-particle interactions; (iv) the Alfv\'en wave amplitude in saturation reflects the expected transfer of net momentum from the originally-anisotropic CRs to forward-propagating Alfv\'en waves.  Here, we adopt the same numerical methods and conduct simulations in a similar parameter regime to \citetalias{bai_19}, but now we additionally  consider the effects of ion-neutral interactions that damp waves, competing with CR-induced wave generation. In separate  work, \citet{bambic_21} have also used MHD-PIC to investigate CR propagation through the multiphase ISM. That study extends \citetalias{bai_19} and the present work by simulating a drifting CR population across a two-component ISM (with and without damping, modeling neutral and ionized regions).

This article is organized as follows. In \autoref{sect:analytics}, we review the analytical derivation  of the CRSI linear growth rate, including the contribution from  ion-neutral damping. In \autoref{sect:numercial_sims} we outline our numerical methods including the implementation of ion-neutral wave damping. Simulation results are presented in \autoref{sect:numerical_results}. We further discuss our findings and their astrophysical context in \autoref{sect:discussion}, and summarize our main conclusions in \autoref{sect:conclusion}.

\section{Cosmic Ray streaming instability including ion-neutral interaction}
\label{sect:analytics}

In this section, we discuss the linear growth rate of the CRSI without and with ion-neutral damping. While analytic studies of CRSI commonly adopt a power-law  distribution, here we will instead  consider a $\kappa$-distribution, as  adopted in our simulations.  This choice is motivated by the use of $\delta f$ method, in which the unperturbed part of the CR distribution must be specified for all $p$, and should be a smooth function \citep[see ][]{bai_19}. Because standard expressions for CRSI with ion-neutral collisions are based on power-law distributions, here we re-derive growth rates for the case of the $\kappa-$distribution. These analytic results are useful for comparison with the simulations that follow.  

Some discussion of the high-frequency regime is given in the last part of this section.

 \subsection{No damping}
 \subsubsection{Resonant contribution only}
 \label{subsubsect:growth_resonant}
We consider the instability arising from resonant interaction in the low-frequency, $\omega \ll \Omega_0$, and non-relativistic drift ($V_D \ll c$) limit\footnote{In this study we adopt the notation $\Omega_0$ for the quantity denoted by $\Omega_c$ in \citetalias{bai_19}, i.e. $\Omega_0 \equiv \Omega_c$.}. Here, $\omega = k V_A$ is the Alfv\'en wave frequency and $\Omega_0 = e B_0 /(m_p c)$ is the non-relativistic proton gyro-frequency.  We adopt standard notation, with Alfv\'en velocity  $V_A= B_0/(4 \pi \rho_i)^{1/2}$ for $B_0$ the local magnetic field strength and $\rho_i$ the background ion density;
 $e$ is the proton charge, $m_p$ is the proton mass, and $c$ is the speed of light. We also assume that the mass and charge of the background fluid ions are identical to those of CR particles\footnote{If this condition is not satisfied, one has to replace $\Omega_0$ by $\Omega_i = q_i B_0 / (m_i c)$ in the final expression of the growth rate, i.e., in Equations \ref{eq:grate_kappa}, \ref{eq:growth_rate_max}, \ref{eq:dispers_bai_19}, \ref{eq:omega_image}, \ref{eq:dispers_general_IN}. We note however, that if mass densities are used instead of number densities then the characteristic frequency appearing in the growth rate would correspond to the gyrofrequency of CRs, owing to the equality $(n_{\rm CR} / n_i) \Omega_i = (\rho_{\rm CR} / \rho_i) \Omega_0$}: $m_i = m_p$ and $q_i = e$.

We start with the growth rate expression as given by \citep[Eq. 69 of Ch. 12]{kulsrud_05} for wavenumber  $k$:
\begin{eqnarray}
\Gamma_{\rm CR}(k) &=& - \pi^2 e^2 \left( \frac{V_A}{c} \right)^2 \left(\frac{V_D}{V_A} -1\right) \times \nonumber \\
&& \int \frac{\partial F}{\partial p} \frac{p_\perp^2}{p} \delta(k p_\parallel - m_p \Omega_0) {\rm d}^3 p \, ,
\label{eq:grate_kulsrud05}
\end{eqnarray}
where $V_D$ is the initial drift speed between the background gas and the frame in which the  cosmic rays have an isotropic distribution (taken along the external magnetic field in the $x$-direction). The $\delta$ function in the integral accounts for the gyro-resonant interaction at the fundamental harmonic, where resonance occurs at $p_\parallel = m_p \Omega_0/k \equiv p_{\rm res}$.

For the (isotropic) distribution $F(p)$ of CRs in the drift frame, we adopt a $\kappa$-distribution, 
\be
F(p) = A \left[ 1+ \frac{1}{\kappa} \left(\frac{p}{p_0}\right)^2 \right]^{-(\kappa+1)} \, .
\label{eq:kappa_def}
\ee
The normalization factor $A$ is given by
\be
A = \frac{n_{\rm CR}}{(\pi \kappa p_0^2)^{3/2}} \frac{\Gamma(\kappa+1)}{\Gamma(\kappa-1/2)} \, ,
\label{eq:normalization_factor}
\ee
where $n_{\rm CR}$ is the number density of cosmic rays, and $\Gamma(x)$ is the Euler $\Gamma$-function.
We generally adopt $\kappa=1.25$, corresponding to $F(p) \propto p^{-4.5}$ for $p \gg p_0$. We note that a $\kappa-$distribution reduces to a Maxwellian in the limit $\kappa \to \infty$. 

\autoref{eq:grate_kulsrud05} becomes
\begin{eqnarray}
&&\Gamma_{\rm CR} = 2 \pi^2 e^2 A \frac{\kappa+1}{\kappa p_0^2}\left( \frac{V_A}{c} \right)^2 \left(\frac{V_D}{V_A} -1\right) \times  \nonumber \\ 
&&\underbrace{\int \left[ 1+ \frac{1}{\kappa} \left(\frac{p}{p_0}\right)^2 \right]^{-(\kappa+2)} \delta(k p_\parallel - m_p \Omega_0) p_\perp^2  {\rm d}^3 p}_I \, .
\end{eqnarray}
With ${\rm d}^3 p={\rm d} \theta {\rm d} p_\parallel p_\perp {\rm d} p_\perp$, the integral $I$ in the previous equation becomes
\be
I =  \frac{2\pi}{k} \int_0^\infty \left[ 1+ \frac{1}{\kappa} \frac{p_{\rm res}^2+p_\perp^2}{p_0^2} \right]^{-(\kappa+2)} p_\perp^3  {\rm d} p_\perp \, .
\ee
Noting that $p_\perp^2=p^2-p_{\rm res}^2$ and that $p_\perp  {\rm d} p_\perp =p  {\rm d} p$, we change the variable of integration from $p_\perp$ to $p$. This leads to
\begin{eqnarray}
I &=&  \frac{2\pi}{k} \int_{p_{\rm res}}^\infty \left[ 1+ \frac{1}{\kappa} \frac{p^2}{p_0^2} \right]^{-(\kappa+2)} p (p^2-p_{\rm res}^2)  {\rm d} p  \nonumber \\
   &=& \frac{\pi}{k} p_0^4 \frac{\kappa}{\kappa+1}\left[ 1+ \frac{1}{\kappa} \frac{p_{\rm res}^2}{p_0^2} \right]^{-\kappa} \, .
\end{eqnarray}
Using $\rho_i= m_p n_i$, $R_{L,0} \equiv p_0/(m_p \Omega_0)$, $p_{\rm res}/p_0=1/(k R_{L,0})$ and inserting the expression from \autoref{eq:normalization_factor} the growth rate can be written as
\begin{eqnarray}
\Gamma_{\rm CR} (k) &=& \frac{\pi^{1/2}}{2 \kappa^{3/2}} \frac{\Gamma(\kappa+1)}{\Gamma(\kappa-1/2)} \frac{n_{\rm CR}}{n_i} \frac{\Omega_0}{k R_{L,0}} \times  \nonumber \\ 
&& \left(\frac{V_D}{V_A} -1\right) \left[ 1+ \frac{1}{\kappa} \frac{1}{(k R_{L,0})^2} \right]^{-\kappa} \, .
\label{eq:grate_kappa}
\end{eqnarray}

The growth rate is maximal at $k R_{L,0}=\sqrt{2-1/\kappa}$. For the adopted value in this study $\kappa=1.25$, it is 
\begin{eqnarray}\label{eq:growth_rate_max}
\Gamma_{\rm max, 0} \simeq 0.28 \Omega_0 \frac{n_{\rm CR}}{n_i}  \left(\frac{V_D}{V_A} -1\right) \, .
\end{eqnarray}
The subscript $``{\rm max, 0}''$ stands for the maximum growth rate of the instability without IN damping. The numerical prefactor varies from $0.25$ for $\kappa=1$ to $0.38$ for $\kappa \to \infty$. In the long-wavelength limit ($k R_{L,0} \ll 1 $) the growth rate varies as $\propto k^{2\kappa -1}$, and in short-wavelength limit ($k R_{L,0} \gg 1 $) as $\propto k^{-1}$.

 \subsubsection{Full dispersion relation}
 
 In more general cases when CR-induced current can be large, the contribution from non-resonant instability can be significant \citep[see, e.g.,][]{bell_04, amato_blasi_09, bykov_13}. The full dispersion relation using the  $\kappa-$distribution was derived in \citetalias{bai_19}. Here, we reproduce it for completeness:
 \begin{eqnarray}
\omega^2 &=&  k^2 V_A^2 \mp \frac{n_{\rm CR}}{n_{\rm i}} \Omega_0 (\omega-k V_D) \left( (1-Q_1) \pm i Q_2 \right) \, ;
\label{eq:dispers_bai_19}
\end{eqnarray}
here the $(1- Q_1)$ and $Q_2$ terms are due to non-resonant and resonant responses of CRs to waves, respectively (see Eq. 9 in \citetalias{bai_19} for definition of these terms). The resonant term given in Eq.~40 of \citetalias{bai_19} is:
\be
Q_2 = \frac{\sqrt{\pi}}{\kappa^{3/2}} \frac{\Gamma(\kappa + 1)}{\Gamma(\kappa - 1/2)} \frac{1}{k R_{L,0}} \left[ 1 + 1/(\kappa k^2 R_{L,0}^2)\right]^{-\kappa} \, .
\label{eq:Q2_term_bai19}
\ee
In the case with negligible contribution from the  non-resonant term ($(1-Q_1)=0$), the imaginary part of $\omega$ obtained by solving the \autoref{eq:dispers_bai_19} is then
\be
\omega_{\mathcal I}= \frac{1}{2} \frac{n_{\rm CR}}{n_i} \Omega_0 \left(\frac{V_D}{V_A} -1\right) Q_2 \, ,
\label{eq:omega_image}
\ee
which is identical to the growth rate given in \autoref{eq:grate_kappa}.

 \subsection{With ion-neutral damping}
 \subsubsection{Damping term from fluid equations}
 \label{subsubsect:Damping_term_from_fluid}
In this section,  we discuss the Alfv\'en wave propagation properties in presence of neutrals, temporarily ignoring the destabilizing contribution from CR streaming.  The derived damping rate is then subtracted from the CRSI growth rate given above to obtain the net growth rate.\footnote{Such a procedure is typically adopted in the literature \citep{kulsrud_cesarsky_71,ginzburg_73,zweibel_shull_82,nava_16}}

Consider coupled two-fluid momentum equations for ions and neutrals, without the contribution of CRs, and where Ohmic dissipation and the Hall effect are neglected:

\begin{eqnarray}
\rho_{\rm i} \frac{\rm{D} \mathbf{u_i}}{\rm{D} t} & = & - \mathbf{\nabla} P_{\rm i} + \frac{\mathbf{J}}{c} \times \mathbf{B} - \rho_{\rm i} \nu_{\rm in} (\mathbf{u_i} - \mathbf{u_n})
\label{eq:momentum_ion}\\
\rho_{\rm n} \frac{\rm{D} \mathbf{u_n}}{\rm{D} t} & = & - \mathbf{\nabla} P_{\rm n} - \rho_{\rm n} \nu_{\rm ni} (\mathbf{u_n} - \mathbf{u_{\rm i}}) 
\label{eq:momentum_neut} \, ,
\end{eqnarray}
where $\rho_{\rm i}$ ($\rho_{\rm n} $), $P_{\rm i}$ ($P_{\rm n} $), $\nu_{\rm in}$($\nu_{\rm ni}$) are the mass density, pressure, and 
collision frequencies of ions with neutral (neutrals with ions), respectively. The quantities $\mathbf{B}$ and $\mathbf{J}$ correspond to the magnetic field and current carried by the plasma. The ions are subject to both the Lorentz force and friction with neutrals, while neutrals are only subject to friction with ions.

From momentum conservation, 
$\rho_{\rm i} \nu_{\rm in} = \rho_{\rm n} \nu_{\rm ni}= n_n n_i \mu \langle \sigma v \rangle$, where $\mu = m_i m_n /(m_i+m_n)$ is the reduced mass, and $\langle \sigma v \rangle$ is the momentum transfer rate coefficient. Most relevant to wave damping is the ion-neutral collision frequency
\begin{equation}
\nu_{\rm in} = \frac{\rho_n}{m_i+ m_n} \langle \sigma v \rangle\ .
\end{equation} 
\cite{draine_11} (see Table 2.1) provides values of the coefficient, $\langle \sigma v \rangle\approx 3.3 \times 10^{-9} \cm^3~\s^{-1}$ for neutral atomic gas where the main collision partner is H with $\mu \approx 0.5 m_p $ for  H$^+$ ions, and $\langle \sigma \upsilon \rangle\approx 1.9 \times 10^{-9} \cm^3~\s^{-1}$ in molecular regions where  the collision partner is H$_2$ with $\mu \approx 2 m_p$ for C$^+$ ions in diffuse gas or HCO$^+$ ions in dense gas.

The magnetic field in 
Equations \ref{eq:momentum_ion}-\ref{eq:momentum_neut} evolves subject to the induction equation:
\begin{eqnarray}
\frac{\partial \mathbf{B}}{\partial t} &=&  \mathbf{\nabla} \times \left( \mathbf{u_i} \times \mathbf{B} \right) \label{eq:induction}
\end{eqnarray}

For simplicity, we consider a static, homogeneous and incompressible background medium. The mean magnetic field is oriented along the $x$-direction, $\mathbf{B}_0 = B_0 \mathbf{e}_x$  and the perturbation components are perpendicular to $\mathbf{B_0}$: $\mathbf{\delta B} = \delta B_y \mathbf{e}_y + \delta B_z \mathbf{e}_z$. The velocity and magnetic field perturbations are  $\propto\exp [i (k x - \omega t)]$. Linearizing Equations \ref{eq:momentum_ion}, \ref{eq:momentum_neut} and \ref{eq:induction} leads for ions to 
\begin{eqnarray}
\omega^2 \mathbf{\delta u_i} &=& k^2 V_{A,i}^2 \mathbf{\delta u_i} - i \omega \nu_{\rm in} (\mathbf{\delta u_i} - \mathbf{\delta u_n} )\, ,
\end{eqnarray}
and for neutrals to 
\begin{eqnarray}
\omega^2 \mathbf{\delta u_n} &=&  - i \omega \nu_{\rm ni} (\mathbf{\delta u_n} - \mathbf{\delta u_i} )\, .
\end{eqnarray}
Here, the Alfv\'en velocity includes the ion density only, i.e., $V_{A,i} = B_0 /\sqrt{4 \pi \rho_{\rm i}}$.
Combining these two equations gives the dispersion relation:
\begin{eqnarray}
\omega^2 \left( 1 +  \frac{i \nu_{\rm in}}{\omega + i \nu_{\rm in}\rho_i/\rho_n
}\right) &=&  k^2 V_{A,i}^2 \, .
\label{eq:IN_two_fluid_linearized}
\end{eqnarray}

One identifies here the Alfv\'en wave dispersion relation (in the ion fluid) modified by the presence of neutrals (second term inside brackets on the left-hand side). \autoref{eq:IN_two_fluid_linearized} is a cubic equation for $\omega$. Despite existing, general solutions are quite cumbersome, hence we do not provide them here. More physical insight can be gained by  considering different limits, previously discussed in the literature \citep[e.g., ][]{kulsrud_pearce_69,zweibel_shull_82,tagger_95, nava_16, xu_2016}. 

Here we reproduce the limits for completeness. We define $Z = \rho_i / \rho_n$ as the ratio of ionized to neutral mass density. It is related to the ionization fraction as $x_i \equiv n_i/n_n = Z/q$, where $q = m_i / m_n$. Three limiting cases are then well defined:
\begin{enumerate}
\item Ion-dominated, $Z \gg 1$. Here, the Alfv\'en wave velocity is unmodified by neutrals, $\omega = k V_{A,i}$. The damping rate is, for any $k$ and $\omega$,
\be
    \Gamma_{\rm d} = \frac{\nu_{\rm in}}{2} \frac{k^2 V_{A,i}^2}{k^2 V_{A,i}^2 + \nu_{\rm in}^2 Z^2}
\ee
For $\omega_k = k V_{A,i} \gg \nu_{\rm ni}=\nu_{\rm in} Z$ it is equal to $\nu_{\rm in}/2$.
\item Neutral-dominated ($Z \ll 1$) and low-frequency ($k V_{A,i}/\nu_{\rm in} \ll \sqrt{Z}$). The fluids are strongly coupled, hence the Alfv\'en wave velocity includes the sum of ion and neutral densities: $\omega = k V_{A, tot} = k B_0 /\sqrt{4 \pi (\rho_{\rm i} + \rho_{\rm n})}$. The damping rate is
\be
    \Gamma_{\rm d} \approx \frac{k^2 V_{A,i}^2}{ 2 \nu_{\rm in}}
\ee
\item Neutral-dominated ($Z \ll 1$) and high-frequency ($k V_{A,i} /\nu_{\rm in} \gg 1$). Ions and neutrals are decoupled. The Alfv\'en wave velocity is unmodified by neutrals, $\omega = k V_{A,i}$. The damping rate is
\be
    \Gamma_{\rm d} \approx \frac{\nu_{\rm in}}{2}
\ee
\end{enumerate}
In neutral-dominated media there is a range of frequencies where there is no propagation (the solution of dispersion relation is purely imaginary): $2 \sqrt{Z} < k V_{A,i} /\nu_{\rm in} < 1/2$. 

\begin{figure}
	\begin{center}
	\includegraphics[width=0.98\columnwidth]{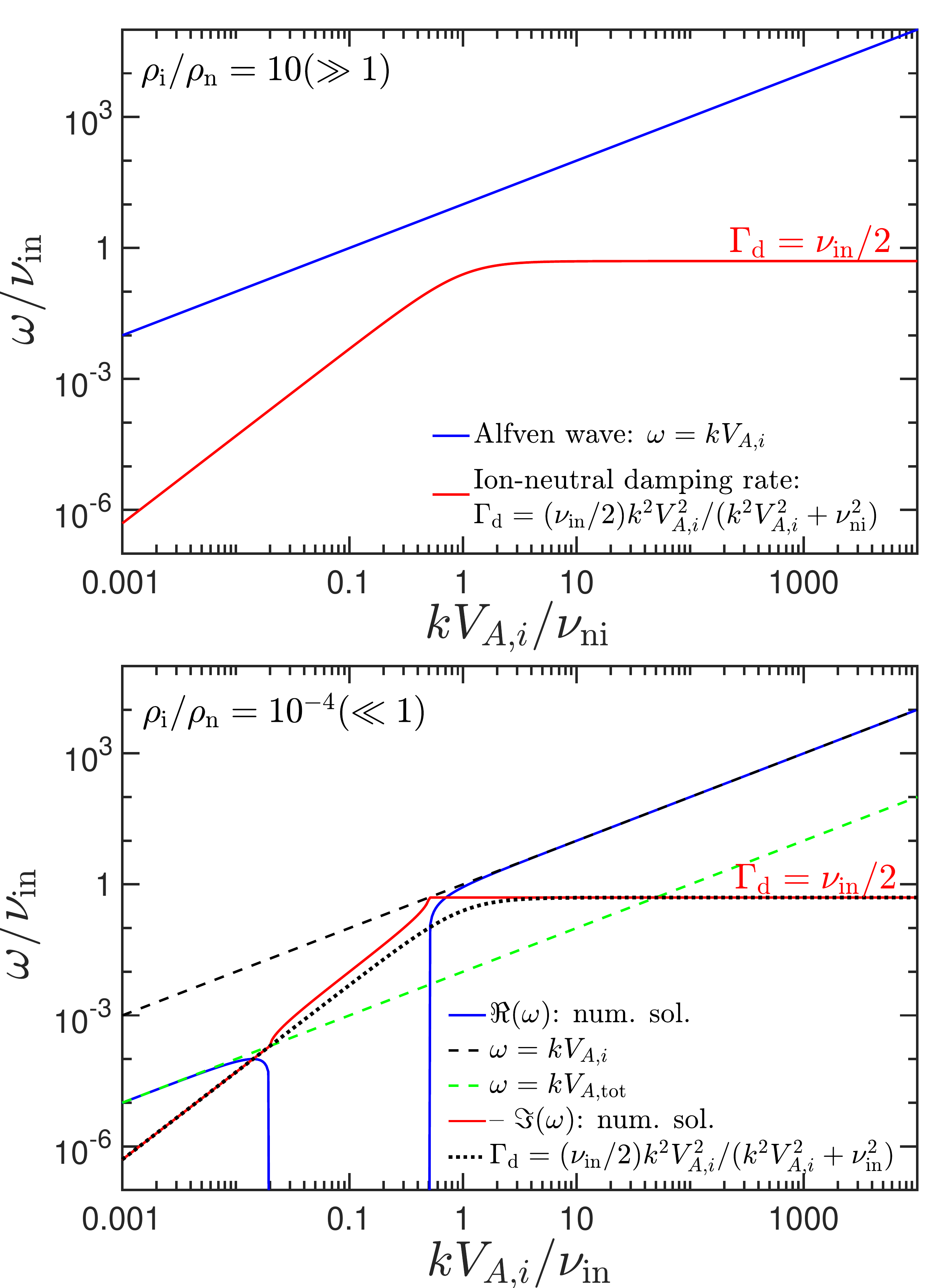}
	\caption{Dispersion relation of Alfv\'en waves and damping rate in presence of neutrals. The two panels present the real and imaginary parts of the wave frequency (blue and red solid lines, respectively) in ion-dominated case (top panel, analytical and numerical solution are identical) and neutral dominated case (bottom panel). The black dotted line in the bottom panel corresponds to an approximate function that fits the damping rate reasonably well  for any $k$ where a solution exists.} \label{fig:AlfvenDispersion_INdamped}
	\end{center}
\end{figure}

\autoref{fig:AlfvenDispersion_INdamped} presents the numerical solutions of \autoref{eq:IN_two_fluid_linearized} and approximate analytical solutions for the ion-dominated case (top panel) and neutral-dominated cases (bottom panel).

In the present study, we will adopt a one-fluid numerical approach for the ionized fluid only. In this case, the dispersion relation (ignoring the dynamics of neutrals) is
\begin{eqnarray}
\omega^2  +  i \nu_{\rm in} \omega &=&  k^2 V_{A,i}^2 \, .
\end{eqnarray}
This implicitly means that we will study the regime where neutrals are fully decoupled from ions, that is the case in high-frequency limit where the Alfv\'en wave is not modified by neutrals $\omega_{\mathcal R} = k V_{A,i}$ and the damping rate is asymptotic at its highest value: $-\omega_{\mathcal I} = \Gamma_d = \nu_{\rm in} / 2$ (see \autoref{fig:AlfvenDispersion_INdamped}). For realistic parameters in the neutral ISM, this limit generally applies (see Section~\ref{sec:highfreq}).

\subsubsection{General dispersion relation with CR streaming contribution}

It is straightforward to generalize the dispersion relation of the CRSI with ion-neutral interactions \citep[see, e.g., ][]{reville_07, reville_21}. Following the procedure of \citetalias{bai_19} but with the addition of ion-neutral momentum exchange terms in fluid equations, we obtain:
\begin{eqnarray}
\omega^2 \left( 1 +  \frac{i \nu_{\rm in}}{\omega + i \nu_{\rm in} \rho_i / \rho_n }\right) &=&  k^2 V_{A,i}^2 \mp \frac{n_{\rm CR}}{n_{\rm i}} \Omega_0 (\omega-k V_D) \nonumber \\
&& \times \left( (1-Q_1) \pm i Q_2 \right) \, ,
\label{eq:dispers_general_IN}
\end{eqnarray}
where $(1-Q_1)$ and $Q_2$ from non-resonant and resonant responses of CRs to waves are discussed above, and $Q_2$ is given for a $\kappa$-DF in \autoref{eq:Q2_term_bai19}.  

Comparing to \autoref{eq:dispers_bai_19}, the supplementary term due to ion-neutral interaction is the second term inside the parentheses on the left-hand side of the \autoref{eq:dispers_general_IN}, while the second term on the right-hand side is due to the CR streaming. For sufficiently small $n_{\rm CR}/n_i$ (as is applicable in general), the solution of this dispersion relation is the same as if the damping (ion-neutral) and growing (CRSI) imaginary terms were derived separately, with the two terms added to obtain the net growth rate. The growth rate from resonant interaction only was presented in the Section~\ref{subsubsect:growth_resonant} and the damping rate in Section~\ref{subsubsect:Damping_term_from_fluid}. Taking the high-frequency limit, the net growth rate is

\begin{eqnarray}
\Gamma_{\rm tot} (k) = \Gamma_{\rm CR} (k) - \frac{ \nu_{\rm in} }{ 2 } \, ,
\label{eq:growth_rate_total}
\end{eqnarray}
where $\Gamma_{\rm CR} (k)$ is given by \autoref{eq:grate_kappa}  for a $\kappa$-DF. This equation is valid for any $k V_{A,i} \gg {\rm max} [\nu_{\rm in}, \nu_{\rm ni}]$. Note that in the low-frequency (long-wavelength) regime, one needs to distinguish between high- and low-ionization fraction cases, which we do not consider here.

\subsection{ISM parameters and implications}
\label{sec:highfreq}

As discussed in  Section~\ref{subsubsect:Damping_term_from_fluid}, the one-fluid formulation does not capture all the details of wave damping. For this reason we discuss here the conditions for which the one-fluid approach is accurate.
 
We assume typical ISM values $B =5\mu$G, $n_{\rm CR}=10^{-9}$~cm$^{-3}$, and typical energy of CRs around $E_0 \sim$ GeV. The frequency of the fastest growing wave mode driven by GeV protons is defined as $\omega_{\rm max} = 1.1 V_{A,i}/ R_{L,0}$, since the fastest growing wavenumber is $k_{\rm max} = 1.1 R_{L,0}^{-1}$ (see text below \autoref{eq:growth_rate_max}). The typical values for the ion-neutral momentum exchange rate $\nu_{\rm in}$ in different phases of the ISM were discussed previously in Section \ref{subsubsect:Damping_term_from_fluid}. 
The high-frequency regime is valid for $\omega_{\rm max} / \nu_{\rm in} \gg 1$. Comparing the two frequencies in different phases of the ISM \citep[as done in \autoref{tab:ISM_phases}, where values for different phases are taken from][Chap. 16]{draine_11} we obtain that this inequality is largely satisfied in all neutral-dominated phases.

Another important consideration concerns the possibility of wave growth in presence of IN damping. The CRSI-driven waves can only grow if the maximum growth rate of the CRSI is larger than the IN damping rate: $\Gamma_{\rm max, 0} / \Gamma_{\rm d} > 1$. Otherwise, even the fastest growing mode is damped. For this purpose, we define a $\zeta$ parameter as
\be
\zeta = \frac{\Gamma_{\rm max, 0}}{ \Gamma_{\rm d}} \frac{1}{V_D / V_{A,i} -1}
\label{eq:zeta_param}
\ee
Note that by dividing \autoref{eq:growth_rate_max} (or more generally \autoref{eq:grate_kulsrud05}) by  the factor  $V_D/V_{A,i}-1$, $\zeta$ is independent of $V_D$. CRSI can be triggered at a range of wavelengths provided that $V_D / V_{A,i}-1 >  1/\zeta$.  
Thus, if $\zeta > 1$, the CRSI can be triggered for super-Alfv\'enic drift velocity $V_D / V_{A,i}-1$ that is order-unity or lower. From \autoref{tab:ISM_phases}, we see that $\zeta$ is typically in the range of 0.1-20 (except in the DMG),
\footnote{For the DMG phase, the required initial drift velocity of CRs is extremely high, $V_D  > 1.5\times 10^{3}  V_{A,i}= 0.3 c$, implying that wave excitation by CRSI is extremely challenging in this medium unless a powerful accelerator is present at close distance, increasing locally the CR density well above $n_{\rm CR} = 10^{-9} \cm^{-3}$.} 
Thus in these environments, and considering only IN damping, the CRSI can be triggered for streaming velocity of $0.1 \lesssim V_D/V_{A,i}-1 \lesssim 10$.  But by the same token, wave damping  will significantly limit the range over which the instability can be directly excited if $V_D/V_{A,i}-1$ is not large compared to $1/\zeta$.

\begin{table}
\caption{List of typical parameter values for different phases of the ISM.}\label{tab:ISM_phases}
\begin{center}
\begin{tabular}{c|cccc|cc}\hline\hline
 Phase & $n$ & $m_i$ & $m_n$ & $x_i$ & $\nu_{\rm in}$ & $\zeta $ \\
   &  ($\cm^{-3}$) & ($m_p$) & ($m_p$) & & ($\omega_{\rm max}$)  & \\\hline
WNM & 0.3 & 1 & 1.4 & $10^{-2}$ & $1.6 \times 10^{-5}$ & 15.5 \\
CNM & 30 & 12 & 1.4 &  $10^{-4}$ & $1.0 \times 10^{-3}$ & 0.07 \\
MG & $10^2$ & 29 & 2.3 & $10^{-6}$  & $3.9 \times 10^{-4}$ & 0.7 \\
DMG & $10^4$ & 29 & 2.3 & $10^{-7}$ & $1.2 \times 10^{-1}$ & $6.6 \times 10^{-4}$ \\
\hline\hline
\end{tabular}
\end{center}
WNM: Warm Neutral Medium, CNM: Cold Neutral Medium, MG: Molecular Gas , DMG: Dense Molecular Gas. Reported quantities are the total density $n$, ion mass $m_i$, ionization fraction $x_i$, corresponding $\omega_{\rm HF}$ above which the high-frequency regime is satisfied, and the drift-normalized instability parameter $\zeta$, that is defined in \autoref{eq:zeta_param}. We adopted $B_0 = 5 \mu$G, $n_{\rm CR} = 10^{-9} \cm^{-3}$, and 
$p_0 = m_p c$ corresponding to CR energy $ E \simeq$GeV, as fiducial values.
\end{table}

\subsection{Wave saturation and steady-state streaming velocity}
\label{subsec:Saturation_KC_criterion_and_QL_scattering}

When CRSI is efficiently triggered, scattering of CRs off of self-generated waves gradually reduces the anisotropy of the CR DF, lowering the drift velocity and the growth rate. The limitation of CR transport by self-generated Alfv\'en waves is commonly referred to as \emph{self-confinement} of CRs. 

In the absence of wave damping, the DF anisotropy (in the wave frame) is eventually erased and the wave amplitude saturates.  Considering that the momentum lost by CRs is transferred to waves, the saturation level can be estimated approximately as \citep[e.g., ][]{kulsrud_05,bai_19, holcomb_spitkovsky_18}

\be
\left(\frac{\delta B}{B_0}\right)^2 \sim \frac{n_{\rm CR}}{n_i}  \left(\frac{V_D}{V_{A,i}} - 1\right) \, .
\label{eq:Bsat_noDamp}
\ee
The wave intensity at saturation is linearly proportional to the number density of CRs and to the first-order anisotropy of the distribution function. 

In the presence of IN damping (or other damping), a commonly adopted assumption is that the streaming velocity is such that the linear growth rate of CRSI balances the wave damping rate \citep[e.g., ][]{kulsrud_cesarsky_71, krumholz_20,hopkins_21a,hopkins_21b}. If we consider the fastest-growing waves and equate to zero the net growth rate with IN damping, $\Gamma_{\rm max, 0}- \nu_{\rm in}/2$, the corresponding streaming velocity (for our fiducial $\kappa=1.25$) is
 \be
\frac{V_{st}}{V_{A,i}} - 1\simeq 1.8 \frac{n_i}{n_{\rm CR}} \frac{\nu_{\rm in} }{\Omega_i} \, .
\label{eq:Streaming_velocity_Gmax}
\ee
Here, the use of $\Omega_i$ instead of $\Omega_0$ in Section~\ref{subsubsect:growth_resonant} is required because the ionized fluids have $m_i > m_p$ in many of the relevant environments (see \autoref{tab:ISM_phases}).

The value in \autoref{eq:Streaming_velocity_Gmax} sets, in principle, a lower limit on CR streaming velocity when IN damping is present. However, even if waves near the fastest-growing wavenumber can grow, waves at much shorter or longer wavelengths would be damped.  As a result, particles with very large (or very  small) momentum that are resonant with smaller (or larger) $k$ would not experience much scattering.   
Thus, the momentum-weighted value of $V_{st}$ will always be larger than the one given in \autoref{eq:Streaming_velocity_Gmax}. 

We note that the conventional approach described above considers only linear growth and damping of waves. In reality, both linear growth \emph{and} pitch angle scattering must be considered, in opposition to wave damping. For a given momentum,  resonant waves  must be excited {\it and} survive IN damping over the timescale required for pitch angle diffusion, in order for isotropization to occur.  Quantifying this requires comparison of the isotropization time (after the linear phase), $1/\nu_s$, with the damping time, $1/\nu_{\rm in}$.

In the absence of damping, the saturation amplitude is given by \autoref{eq:Bsat_noDamp}, and since the scattering rate under QLD is $\nu_s \sim (\pi/8) \Omega (\delta B/B_0)^2$ \citep{kulsrud_05}, comparing to \autoref{eq:growth_rate_max} shows that $\nu_s  \sim \Gamma_{\rm max, 0}$. If we were to assume that full redistribution in $\mu$ requires a time $\sim \nu_s^{-1}$, the total time for isotropization (considering wave growth and scattering) would be at least $2\Gamma_{\rm max, 0}^{-1}$; in practice the numerical results described below (see \autoref{fig:Global_evol_nuINs}) correspond to a prefactor $\sim 10$ for the isotropization time.  With IN  damping, the growth rate is reduced, and (as we shall show) the maximum level of $(\delta B/B_0)^2$ is also reduced, which slows scattering.  
One might therefore expect that for isotropization  to be successful, the ratio $\Gamma_{\rm max, 0}/(\nu_{\rm in}/2)$ must be above some critical value which is larger (perhaps much  larger) than 1, in order to allow for both wave growth and particle diffusion. We return to this issue in \autoref{subsect:non_linear}. 

%%%%%%%%%%%%%%%%%%%%%%%%%%%%%%%%%%%%%%%%%%%%%%%%%%%%%%%%%%%%%%%%%%
% New section
%%%%%%%%%%%%%%%%%%%%%%%%%%%%%%%%%%%%%%%%%%%%%%%%%%%%%%%%%%%%%%%%%%
\section{Numerical simulations using MHD-PIC approach}
\label{sect:numercial_sims}

We use the MHD-PIC method introduced by \citet{bai_15} and adopted for the study of CR-streaming in \citetalias{bai_19}. The focus of the present study is the inclusion of ion-neutral wave damping. Let us recall key  elements of this method:
\begin{itemize}
\item The background thermal plasma dynamics is governed by the ideal MHD equations without explicit viscosity or resistivity, adopting an adiabatic equation of state with $\gamma=5/3$ and initial sound speed $c_s = V_{A,i}$.   The momentum update includes an explicit damping term, implementation of which is described in Section \ref{sec:damping_treatment}. The dynamics of CRs are solved by a PIC method, where CRs are treated as charged macro-particles. The Lorentz equation is solved using a relativistic Boris pusher, the density and electric current are deposited on grid using 2nd order shape factors (TSC scheme), and the coupling between CRs and the background fluid is achieved by introducing CR source terms into the MHD equations and modifying Ohm's law.
\item The  $\delta f$ method is employed to significantly reduce the Poisson noise of macro-particles.
\item Phase scrambling is applied when particles cross the system boundary and re-enter on the other side of the periodic box; this effectively mimics a larger numerical box.
\item Seed waves are initialized at $t=0$. This provides better control of $k$-by-$k$ growth rate as compared to growth from numerical noise (typically done in full-PIC simulations).
\end{itemize}

\subsection{Numerical treatment of the ion-neutral damping}\label{sec:damping_treatment}

The presence of neutrals introduces momentum exchange between ions and neutrals through collisions. The corresponding terms in the fluid equations were given previously in \autoref{eq:momentum_ion} and \autoref{eq:momentum_neut}. Instead of solving the coupled system of two-fluid equations, we only account for the effect of neutrals on the ions while the dynamical equations for neutrals are not evolved. This procedure is consistent in the limit where the two fluids are decoupled (high-frequency regime). The procedure is as follows. At each time step we reduce
the transverse momentum fluctuations according to \autoref{eq:momentum_ion}. After each numerical time step $\Delta t$ the transverse momentum is updated as $p_{\rm \perp, new} = p_\perp \exp(-\nu_{\rm in} \Delta t)$, in order to account for the effect of damping. While the longitudinal velocity should be subject to the same ``damping", it is not incorporated as it is decoupled from the Alf\'ven waves (which have no longitudinal motion) and we have verified that it does not affect the overall simulation results.

\subsection{Numerical setup}

In the present work, there  are  some differences compared to the \citetalias{bai_19} fiducial setup. Namely: \\
(i) We use fewer CR macro-particles per cell, as the  $\delta f$ method  effectively controls the Poissonian noise level. \\
(ii) The physical size of the grid is slightly smaller. Yet, the box size is equal to $\simeq 47$ times the most unstable wavelengths. \\
(iii) We include the wave damping term in the MHD momentum equation.\\
(iv) The fiducial drift velocity is $10 V_{A,i}$ (instead of $2 V_{A,i}$ used in \citetalias{bai_19}). This choice is motivated by the need to efficiently reach a fully saturated state when there is no ion-neutral damping. This case will be used as reference when measuring the effect of different levels of damping.\\
 
\begin{table*}
\caption{List of main simulation runs}\label{tab:params}
\begin{center}
\begin{tabular}{c|ccc|cccc|c}\hline\hline
 Run & $V_D/V_{A,i}$ & $n_{\rm CR}/n_{\rm i}$ & $\nu_{\rm in}/\Gamma_{\rm max, 0}$ & Domain size & Domain size & resolution &  $N_p$ &  Runtime  \\
   & & & & $L_x$ ($d_i$) & $L_x/\lambda_m$ &
$\Delta  x$ ($d_i$)   
&  (per cell) & ($\Omega_0^{-1}$) \\\hline
Fid & 10.0 & $1.0\times10^{-4}$ & $0$ & $8\times10^4$ & 42.4 & $10$ & $ 128 $ & $10^6$ \\
Fid-Damp & 10.0 & $1.0\times10^{-4}$ & $\in [0.03, 1.9]$ & $8\times10^4$ & 42.4 & $10$ & $128$ & $10^6$ \\
HiRes & 10.0 & $1.0\times10^{-4}$ & $\{0, 0.1, 0.5\}$ & $8\times10^4$ & 42.4 & $2.5$ & $128$ & $<10^6$ \\
\hline\hline
\end{tabular}
\end{center}
%\begin{center}
Fixed parameters: ${\mathbb C}/V_{A,i}=300$, $p_0/(m V_{A,i})=300$, $\kappa=1.25$,
and initial wave amplitude $A=10^{-4}$. 
In all models the most unstable wavelength is $0.91 \lambda_m$ for
$\lambda_m=2\pi p_0/(m\Omega_0)\approx 1885 d_i$. The ion skin depth $d_i \equiv {\mathbb C}/\omega_i = V_{A,i}/\Omega_0$. The quantity $\Gamma_{\rm max, 0}$ is defined in \autoref{eq:growth_rate_max} as the maximum growth rate of the CRSI without IN damping (i.e. $\nu_{\rm in} = 0$).
%\end{center}
\end{table*}

\autoref{tab:params}  summarizes numerical parameters and model values adopted in our simulations. Our fiducial choice is $n_\mathrm{CR}/n_i =10^{-4}$.  This is somewhat larger than realistic values in the neutral ISM ($n_\mathrm{CR}/n_i \sim 3 \times 10^{-7} - 10^{-5}$, see \autoref{tab:ISM_phases}), for numerical  expediency; lower  values would have very low CRSI growth rate (requiring very long simulation duration) and low saturation amplitudes (exacerbating the numerical issue of crossing $\mu=0$ for practical resolution). This  choice does not affect any of our theoretical conclusions. The box length is chosen to be much larger than the fastest growing wavelength of the instability. Lengths are given in units of ion skin depth $d_i \equiv {\mathbb C}/\omega_{p,i} = V_{A,i}/\Omega_0$, which is 1 in code units. We adopt a reduced speed of light ${\mathbb C} = 300 V_{A,i}$.  Seed waves are initialized with equal amplitudes of right-handed  and left-handed polarization and the same amplitude at all $k$.  The initial total power in the waves is $\delta B/B_0=10^{-4}$.  We work in the frame where CRs are initially at rest, and gas moves in the $-\mathbf{e}_x$  direction. 

It is convenient to parameterize damping rates $\nu_{\rm in}$ relative to the peak growth rate for CRSI; this is what is reported in \autoref{tab:params}. Using  $V_D=10 V_{A,i}$ and $n_\mathrm{CR}/n_i =10^{-4}$ in \autoref{eq:growth_rate_max}, this initial peak growth rate (for $\nu_{\rm in}=0$) is 
$\Gamma_{\rm max, 0}/\Omega_0 = 2.5 \times 10^{-4}$.

%%%%%%%%%%%%%%%%%%%%%%%%%%%%%%%%%%%%%%%%%%%%%%%%%%%%%%%%%%%%%%%%%%
% New section
%%%%%%%%%%%%%%%%%%%%%%%%%%%%%%%%%%%%%%%%%%%%%%%%%%%%%%%%%%%%%%%%%%
\section{Simulation results}
\label{sect:numerical_results}

In this section we describe the results of MHD-PIC simulations. We start with a presentation of the overall time-evolution, then show the linear growth rates of the instability.  We then discuss the transition into saturated phase and some aspects of the saturated state of the streaming CR-background fluid system. 
 
 \subsection{Overall evolution}
  \label{subsect:overall}
In this section we compare the evolution between no-damping and damping cases (models Fid and Fid-damp reported in \autoref{tab:params}). For the cases with damping we explored values of $\nu_{\rm in}$ between $0.03 \Gamma_{\rm max, 0}$ and $1.9 \Gamma_{\rm max, 0}$; the theoretical critical value for no growth is  $\nu_{\rm in} = 2\Gamma_{\rm max, 0}$. 
In the following, we use the term ``moderate damping" for cases where $\nu_{\rm in} / \Gamma_{\rm max, 0}$ is non-negligible but still smaller than unity. The representative case is $\nu_{\rm in} / \Gamma_{\rm max, 0} = 0.5$.
 
\begin{figure}
\begin{center}
\includegraphics[width=0.45\textwidth]{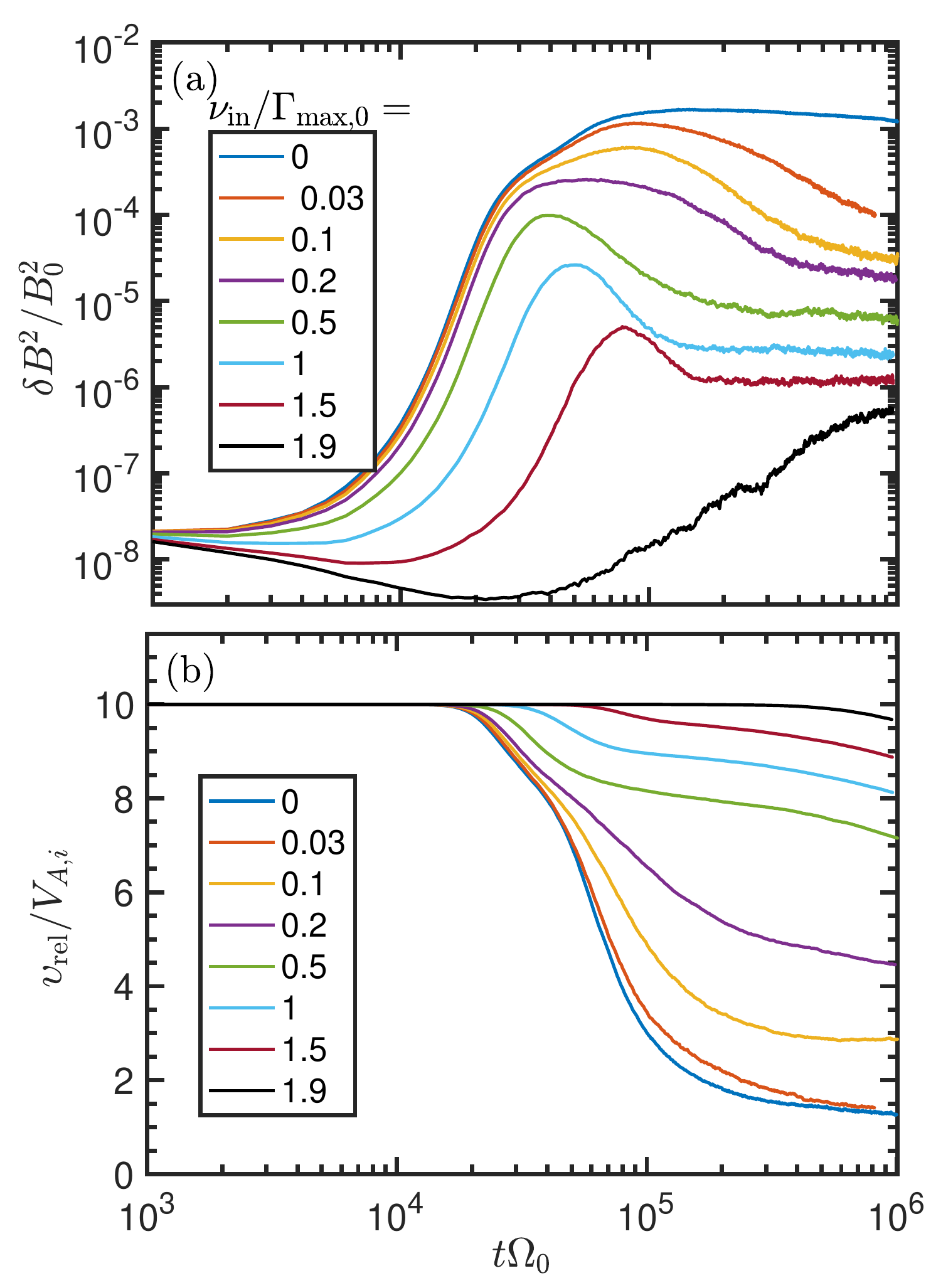}
\end{center}
\caption{Time evolution of the magnetic wave energy (top panel) and of the streaming velocity of bulk CRs (bottom panel) for different values of $\nu_{\rm in}$ ranging from $0$ (blue line) to $1.9 \Gamma_{\rm max, 0}$ (black line).} \label{fig:Global_evol_nuINs}
\end{figure}

\autoref{fig:Global_evol_nuINs} presents the time-evolution of the magnetic wave energy (upper panel) and of the relative velocity between the bulk CRs and the background fluid (bottom panel) for different values of $\nu_{\rm in}$.
The undamped case (blue curves) reaches saturation at $t\Omega_0 \simeq 10^5$ with $\delta B^2/B_0^2 \simeq 10^{-3}$.  The CR streaming velocity, initially $V_D=10 V_{A,i}$, begins to  decrease when $\delta B^2/B_0^2 \simeq 10^{-4}$ and reaches $v_{\rm rel}\simeq V_{A,i}$ by the end of the simulation. 
Here, the CR bulk velocity is defined (in the simulation frame) by 
\be
v_{\rm rel} \equiv \frac{\int f(p,\mu) v(p) p^2 \mu \rm{d} \mu \rm{d} p}{\int f(p,\mu) p^2 \rm{d} \mu \rm{d} p} - V_{bg}    \, ,
\label{eq:streaming_velocity_def}
\ee 
where $V_{bg}$ the background gas speed.
Unlike the initial conditions, the CR distribution is not perfectly isotropic in the frame moving at $v_{\rm rel}$.

Cases with non-zero damping reach lower saturation levels, with the maximum in $\delta B^2/B_0^2$ decreasing as $\nu_{\rm in}$ increases. After saturation (at late simulation times, $t\Omega_0 > 10^5$) the wave intensity stabilizes at intensity that decreases with increasing $\nu_{\rm in}$ \footnote{We noticed that the level of the wave intensity `plateau' at the late-time state can evolve somewhat differently depending on numerical resolution and number of particles per cell, but does not affect qualitatively the system evolution.}.  Streaming velocities (bottom panel) decline at lower rates for the stronger-damping models, because the lower wave amplitudes scatter CRs less effectively. The late-time (after $t\Omega_0 \sim 10^6$) evolution is asymptotically slow.

In general, we define three representative phases of time-evolution of the instability: linear, post-linear, and saturated (late-time). There is an adjustment phase between the time when the fastest modes stop growing (end of linear phase) and the saturation of the instability  when the anisotropy of the distribution function is (eventually) erased (start of saturated phase). We refer to this transitory phase as \emph{post-linear}.

\begin{figure}
\begin{center}
\includegraphics[width=0.47\textwidth]{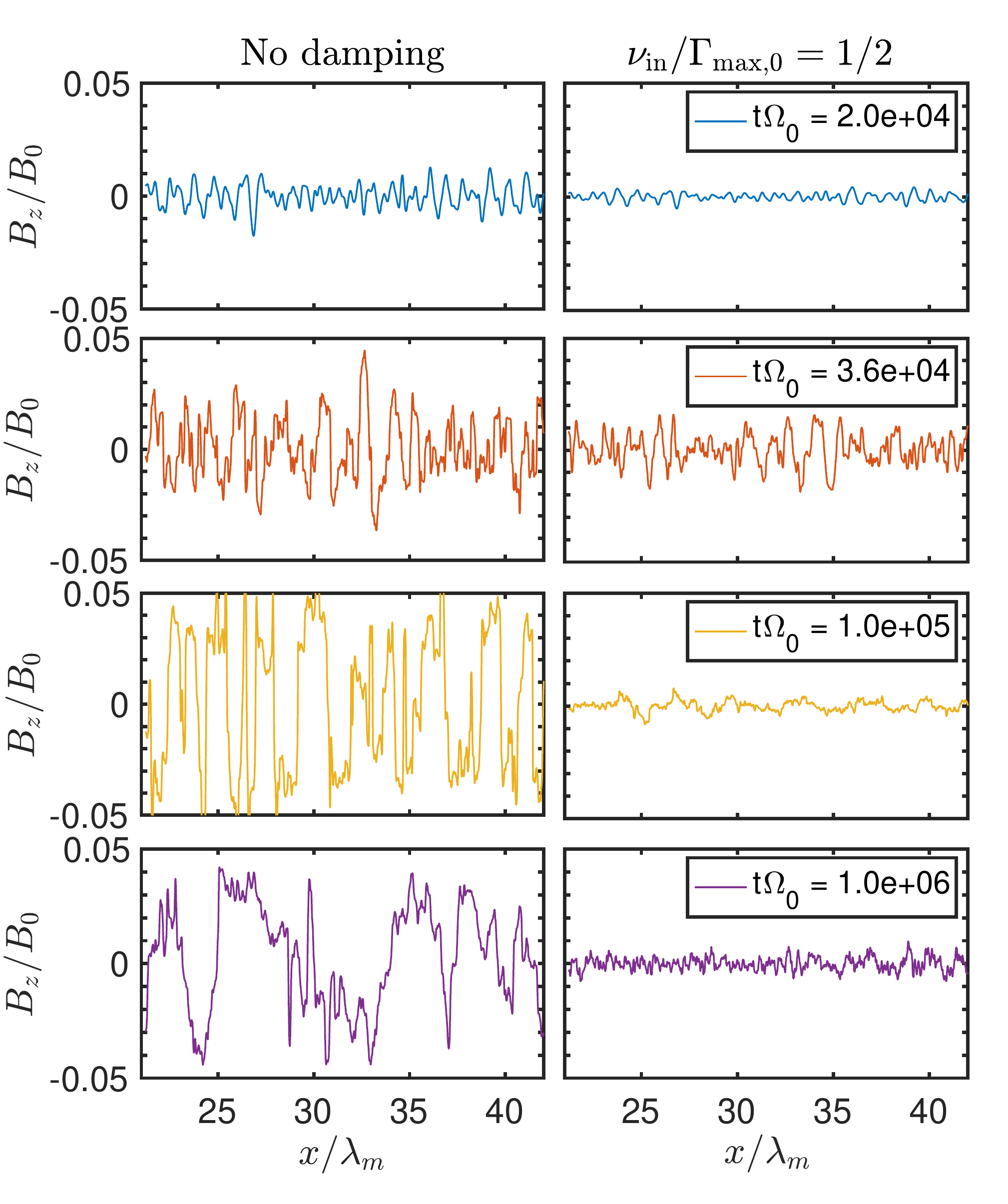}
\end{center}
\caption{Profile of the wave component $B_z$ at different simulation times for the case with no damping (left column) and with moderate damping $\nu_{\rm in} = (1/2) \Gamma_{\rm max, 0}$ (right column). Four different times are presented. The top set corresponds to the early linear phase. The two middle panel sets show the evolution during post-linear phase, and the bottom set shows the final simulation time $t \Omega_0 =10^6$.} \label{fig:Bz_multitime}
\end{figure}

In \autoref{fig:Bz_multitime} we show the magnetic field profile $B_z$ at four different simulation times for the case with no damping (left column) and moderate damping (right column). For the case with no damping the wave amplitude grows to $\delta B_z/B_0 \simeq 0.02$ at the end of the linear phase ($t \Omega_0  = 2 \times 10^4$) and further increases to $0.05$ during the post-linear evolution (between $t \Omega_0 \simeq 3 \times 10^4$ and $t \Omega_0 = 10^5$), during which amplitudes increase at $k$ below the peak (see Section~\ref{subsect:post_linear}). For the case with moderate damping, the wave amplitude is reduced, as expected. During the post-linear phase we observe a decrease of the wave amplitude from $\delta B_z/B_0 \simeq 0.02$ to $4 \times 10^{-3}$, while  there is less contribution from $k$ below the peak.

\begin{figure}
\begin{center}
\includegraphics[width=0.48\textwidth]{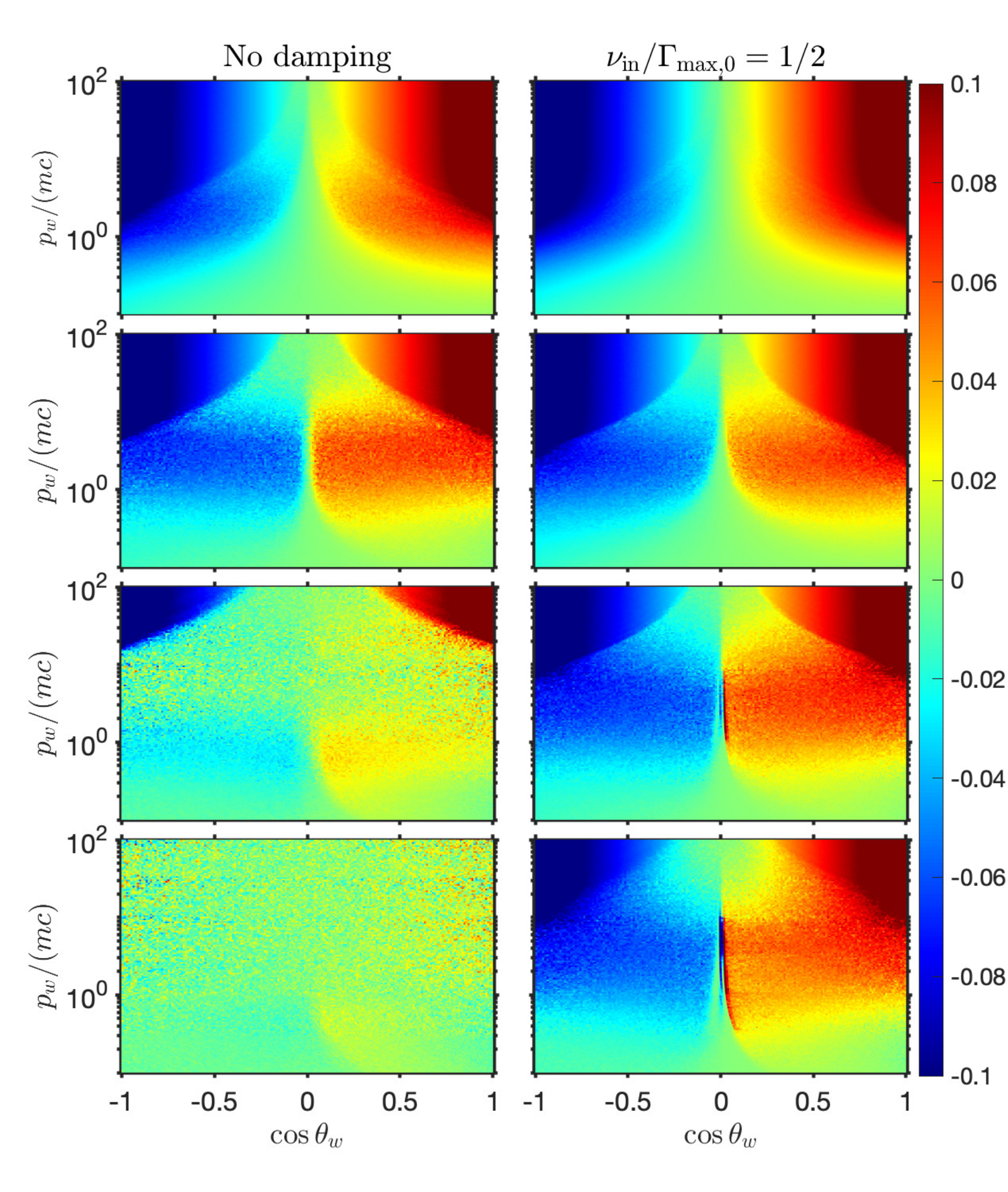}
\end{center}
\caption{2D distribution function in the wave  frame, $\delta f_w(p_w, \cos \theta_w) / f_0$, at different simulation snapshots for the case with no damping (left column) and with moderate damping $\nu_{\rm in} = (1/2) \Gamma_{\rm max, 0}$ (right column). The  same simulation times are presented as in \autoref{fig:Bz_multitime} from top to bottom.} \label{fig:deltaF_multitime_1}
\end{figure}

The effect of CRSI and subsequent QLD on the CR distribution function is presented in \autoref{fig:deltaF_multitime_1}. We show the distribution function in the wave frame (i.e. moving to the right at $V_{\rm A,i}$ with respect to the gas, see \citetalias{bai_19} for its transformation from simulation frame), $\delta f_w(p_w, \cos \theta_w) / f_0$  for the case with no damping (left column) and with moderate damping (right column), at the same simulation times as in \autoref{fig:Bz_multitime}. By inspecting the left column, we observe the gradual suppression of the initial anisotropy with time (evolution from top to bottom). The anisotropy is globally maintained  for the case with damping, although at moderate $p_w$ the distribution becomes relatively flat on each side of $\mu_w=0$. 
The effect of particle accumulation at $\mu_w = \cos \theta_w = 0$ is also evident in the two bottom panels on the right.

 \subsection{Linear phase}
  \label{subsect:linear}
  
 \subsubsection{No damping vs moderate damping}
 
\begin{figure}
\centering
\includegraphics[width=0.45\textwidth]{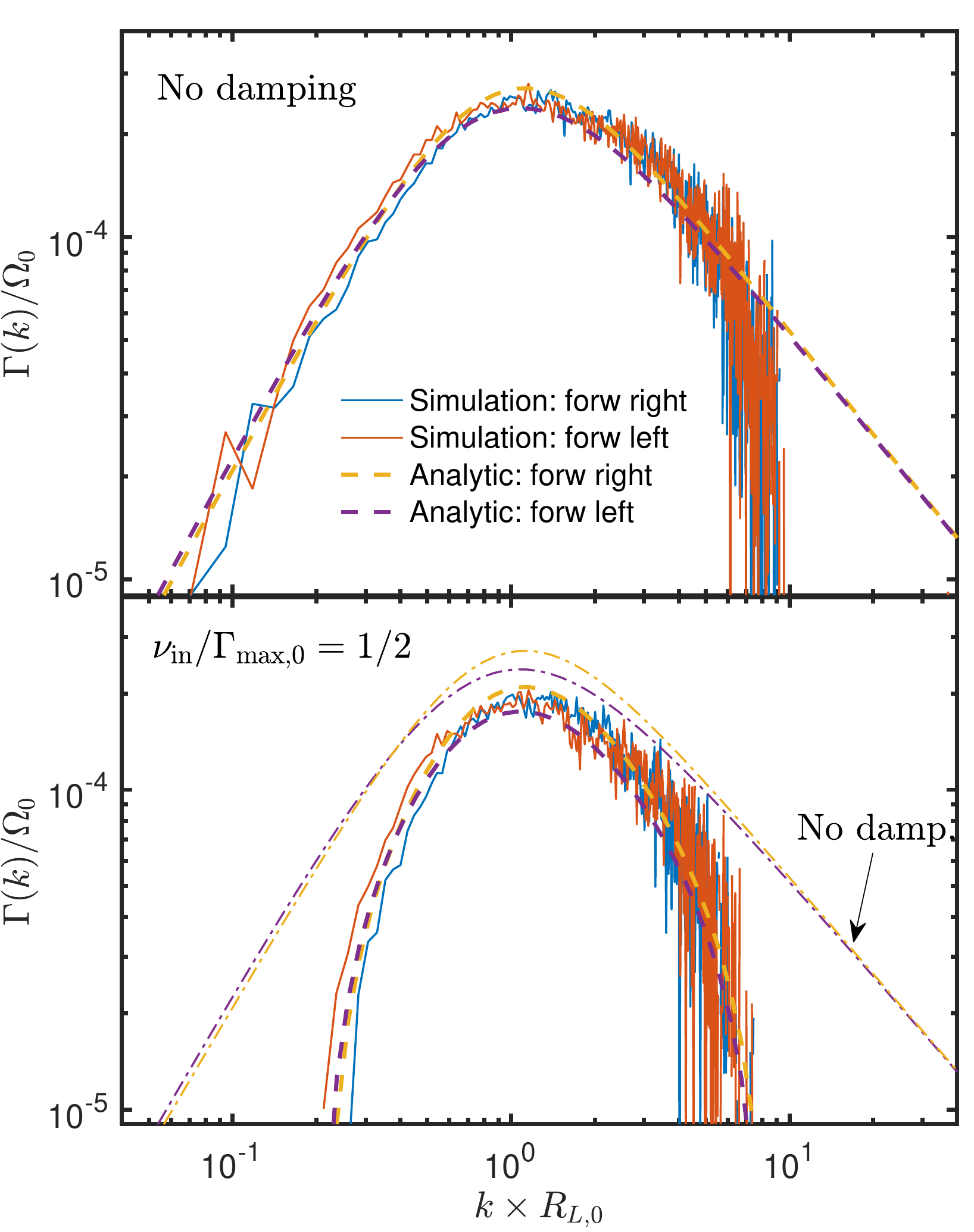}
\caption{Linear growth rate as function of the wavenumber $k$ for the case without damping (top panel) and with moderate damping (bottom panel). Blue and red lines correspond to the measured growth rate of right and left handed modes, respectively. Dashed yellow and magenta lines correspond to the analytical expectation from \autoref{eq:growth_rate_total}. In the bottom panel the analytical expectations for zero damping are also plotted (dot-dashed) for comparison.} \label{fig:linear_grate_damp_nodamp}
\end{figure}

\citetalias{bai_19} demonstrated that our MHD-PIC numerical approach accurately reproduces  the linear growth rate of the CRSI in the absence of explicit wave damping.  Here, we extend the previous investigation by including ion-neutral damping. In \autoref{fig:linear_grate_damp_nodamp}, we show the linear growth rate $\Gamma_{\rm tot}$ as function of the wavenumber $k$ for two cases: no damping (top panel) and moderate damping (bottom panel). 
There is very good agreement with theory, despite some stochastic noise. In particular, the most unstable mode is very well captured at $k_{\rm max} = \sqrt{2-1/\kappa} m _i\Omega_0 / p_0 \simeq 1.1 R_{L,0}^{-1}$.

Comparing the top and bottom panels of \autoref{fig:linear_grate_damp_nodamp}, two immediate effects due to the IN damping can be identified. The first is the decrease of the maximum growth rate near $k R_{L,0} = 1$. It is equal to $2.5\times 10^{-4} \Omega_0$ without damping, and equal to $\simeq 1.8 \times 10^{-4} \Omega_0$ when $\nu_{\rm in} = 1.25 \times 10^{-4} \Omega_0 = \Gamma_{\rm max, 0} / 2$. The second effect is the suppression of wave growth in the low-end and high-end parts of the spectrum, where $\Gamma_0(k) - \nu_{\rm in} / 2 < 0$. In other words, the bandwidth of linearly growing waves is reduced with increasing $\nu_{\rm in}$. 

\subsubsection{Dependence on $\nu_{\rm in}$}

\begin{figure}
\begin{center}
\includegraphics[width=0.45\textwidth]{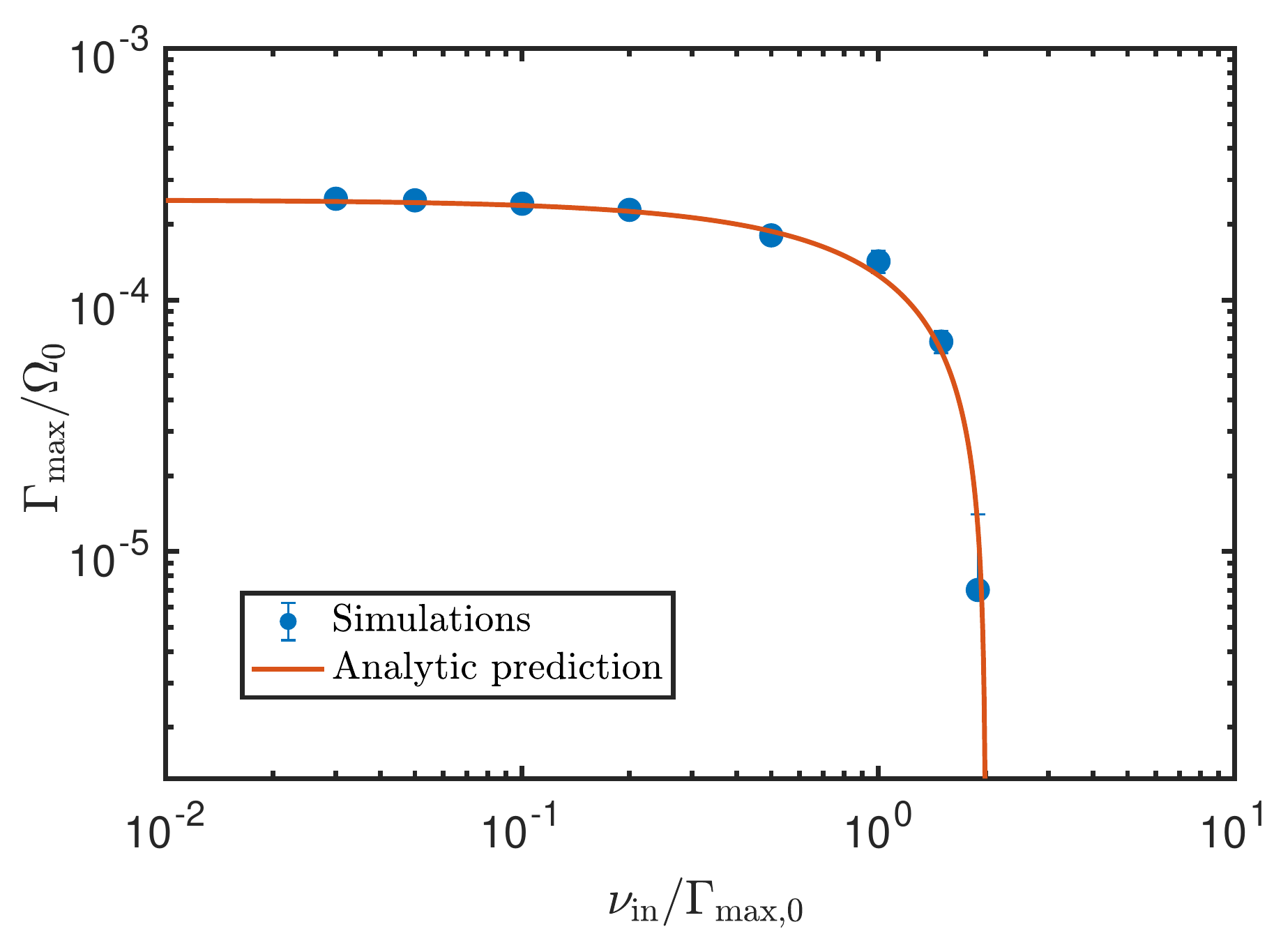}
\end{center}
\caption{Dependence of the maximum linear growth rate of the instability as function of the ion-neutral momentum exchange rate $\nu_{\rm in}$ for $n_{\rm CR}/n_{\rm i}=10^{-4}$ and $V_D/V_{A,i} = 10$.} \label{fig:linear_Gmax_nuIN_dependence}
\end{figure}

In \autoref{fig:linear_Gmax_nuIN_dependence} we report the dependence of the maximum growth rate of the instability on $\nu_{\rm in}$.
  Simulation results are generally in excellent agreement with analytical predictions, namely, the linear growth rate is only weakly reduced for $\nu_{\rm in}/\Gamma_{\rm max, 0} < 0.2$, and rapidly decreases for $\nu_{\rm in}/\Gamma_{\rm max, 0} > 0.5$. As expected from \autoref{eq:growth_rate_total}, there can be no linear growth for $\nu_{\rm in}/\Gamma_{\rm max, 0} > 2$, as even the fastest growing mode is damped.
  
 \subsection{From linear to saturated phase}
 \label{subsect:post_linear}
 
 The instability does not transition directly from linear growth to a fully saturated state. At some point in time the fastest growing modes at $k R_{L,0} \simeq 1$ cease exponential growth. However, at that time the other modes with smaller growth rate can continue to grow because the anisotropy in the distribution function -- which drives all modes -- is not erased. At the same time, wave-wave interaction can redistribute some wave energy from $k R_{L,0} \simeq 1$ to other wave modes. 

The post-linear phase can be easily identified in \autoref{fig:Global_evol_nuINs} for $\nu_{\rm in}=0$ and $\nu_{\rm in} \leq 0.1$. The initial exponential growth slows down roughly at $t\Omega_0 \simeq 3\times 10^4$ but additional wave growth continues until $t\Omega_0 \sim 10^5$. This phase also corresponds to the fastest rate of decrease in the streaming velocity of CRs, as seen in the lower panel of \autoref{fig:Global_evol_nuINs}.

 \begin{figure}
\begin{center}
\includegraphics[width=0.45\textwidth]{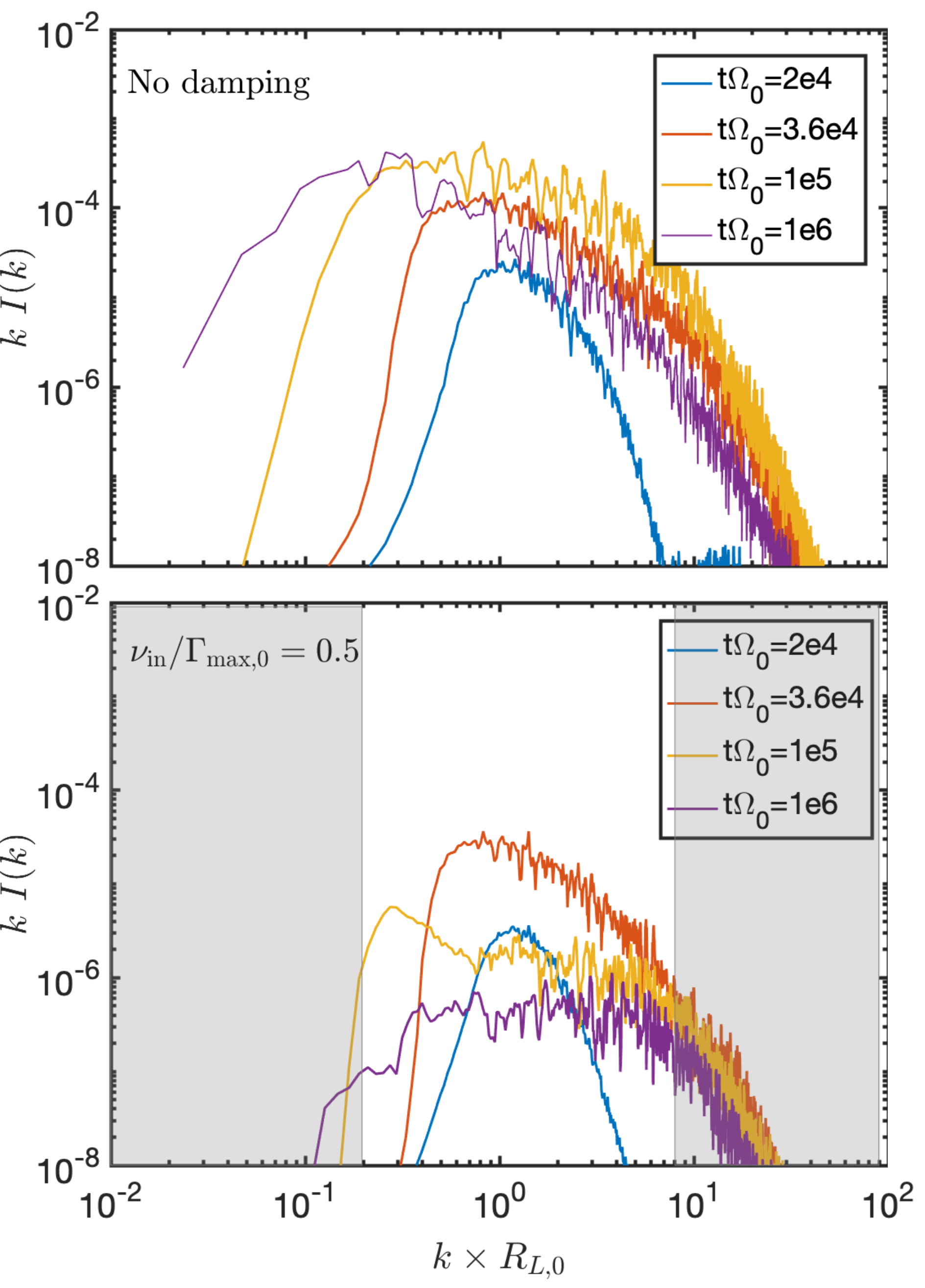}
\end{center}
\caption{Spectra of forward right-handed modes at different times for the case with no damping (top panel) and moderate damping (bottom panel). Evolution for forward left-handed modes is identical. In the lower panel, the grey region marks where initial linear growth is suppressed by IN damping.}
\label{fig:spectra_twocase_lin_to_nonlin}
\end{figure}

More detailed insight into the post-linear phase can be gained by considering the wave spectrum. In \autoref{fig:spectra_twocase_lin_to_nonlin} we present the evolution of the wave spectrum for the case with no damping (top panel) and with $\nu_{\rm in} / \Gamma_{\rm max, 0} = 0.5$ (bottom panel). We note the spectrum is narrow at the end of the linear phase, as shown by the blue line. It is peaked at $k R_{L,0} \simeq 1$, where the growth rate is the fastest. During the post-linear phase (blue to red line), while the growth at $k R_{L,0} \simeq 1$ becomes slow, the modes at $k R_{L,0} > 1$ start growing at the rate comparable to $\Gamma_{\rm max}$, while the modes with $k R_{L,0} \leq 1$ continue to grow approximately at the rate expected from the linear calculation. If there is no damping, during the late evolution the peak in the wave spectrum gradually shifts to larger wavelengths (smaller $k$) while the global level of wave intensity is roughly unchanged. This effect can be clearly seen by comparing red, yellow and purple lines in the top panel: the low-$k$ cut-off shifts from $k R_{L,0} \simeq 0.3$ at the end of the linear phase to $k R_{L,0} \simeq 0.02$ at the end of the simulation. The growth of high-$k$ modes (i.e., $k R_{L,0} > 5$) is observed only during the post linear phase, while the increase of spectral energy in $k R_{L,0} \ll 1$ modes continues well beyond, into the saturated phase.

For the case with  moderate damping ($\nu_{\rm in} / \Gamma_{\rm max, 0} = 0.5$),  the early evolution of the spectrum follows the same trend as in the case with  no damping (blue and red lines in bottom panel). Interestingly, there is a similar fast rise of modes at $k R_{L,0}  \geq 1$ in the post-linear phase, despite the fact that some of these modes are not supposed to grow linearly (there would be a cut-off for $k R_{L,0} > 8$, marked in grey in the figure).
The late-time evolution of the damped case is significantly different from the undamped case. There is a noticeable overall decrease in the wave intensity at all $k$ (difference between red, yellow and purple lines). The peak of the spectrum shifts slightly towards smaller $k$ at the end of the post-linear phase. Also, a spectral bump appears at $t\Omega_0 = 10^5$ around $k R_{L,0} = 0.2$, which is due to driving from the part of the CR distribution function that remains strongly anisotropic: $f(p > 5 p_0)$. By the end of the simulation the spectrum stabilizes. It is flat and narrow: $0.1 < k R_{L,0} <10$ and $k I(k) \sim 2 \times 10^{-7}$. Simulations with different values of $\nu_{\rm in}$ follow similar time-evolution. 

 \subsubsection{Growth of high-$k$ (small wavelength) modes: HiRes simulations}

An important aspect of the post-linear phase is the rapid growth of modes with $k R_{L,0} \geq 5$, \emph{even when these modes are not unstable to CRSI because of ion-neutral damping}. To better study these short-wavelength modes we have conducted additional simulations with higher numerical resolution: $\Delta x =  2.5 d_i$ in the HiRes runs instead of  $\Delta x= 10 d_i$ in Fid simulations. To understand the mechanism driving this growth, we also performed a numerical experiment where the CRs were  removed from the system after a given simulation time, at the end of the linear phase.

\begin{figure}
\begin{center}
\includegraphics[width=0.45\textwidth]{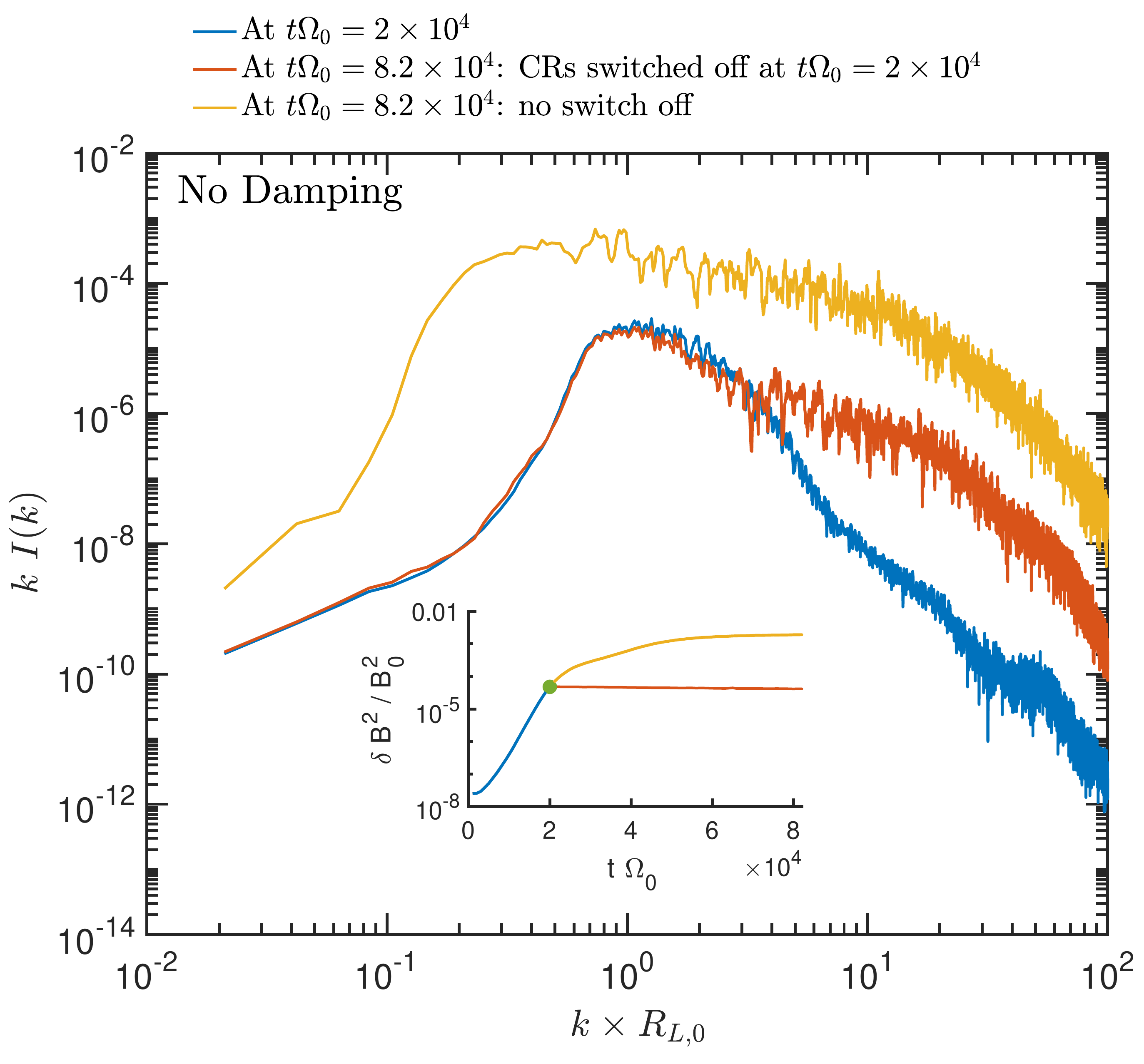}
\caption{Wave spectrum (forward right-handed mode) for the HiRes simulation with no damping at different simulation times. Blue and yellow lines show the standard CR-driven system at $t\Omega_0 = 2\times 10^{4}$ and $t\Omega_0 = 8.2\times 10^{4}$, respectively.  The red line shows the case of a model where CRs are turned off at the end of the linear phase  ($t\Omega_0 = 2\times 10^{4}$) and then freely evolves until  $t\Omega_0= 8.2\times 10^{4}$. The difference between blue and red lines illustrates the growth of high-$k$ modes during the post-linear phase without driving by CRs. The inset in the bottom  shows the time evolution of the magnetic wave energy: the blue line follows the standard evolution until $t\Omega_0 = 2\times 10^{4}$, the yellow line continues the evolution with CRs after that time, and the red line follows the case where CRs are switched off at $t\Omega_0 = 2\times 10^{4}$. The green dot marks the time of the CRs switch-off.} \label{fig:spectrum_swith_off}
\end{center}
\end{figure}

In \autoref{fig:spectrum_swith_off} we show the wave spectrum of forward right-handed modes in the HiRes simulation with no damping. Three lines are plotted: (i) the spectrum at the end of the linear phase (blue line), (ii) the spectrum at the end of the post-linear phase if CRs are switched off at the end of the linear phase (red line), and (iii) the spectrum at the end of the linear phase in the standard case with no CRs switch off (yellow line).  In the inset, we see that wave energy stays constant after switching off the CRs, while continue to grow otherwise. The difference between blue and red lines illustrates the growth of high-$k$ modes during the post-linear phase \textit{without} driving by CRs. This difference is only seen in the region $k R_{L,0} \geq 5$. For comparison, the difference between the blue and yellow lines shows the post linear evolution when the driving by CRs is maintained. Here, the overall increase of the wave energy is observed, together with the shift of the spectrum to lower $k$.

\begin{figure}
\begin{center}
\includegraphics[width=0.45\textwidth]{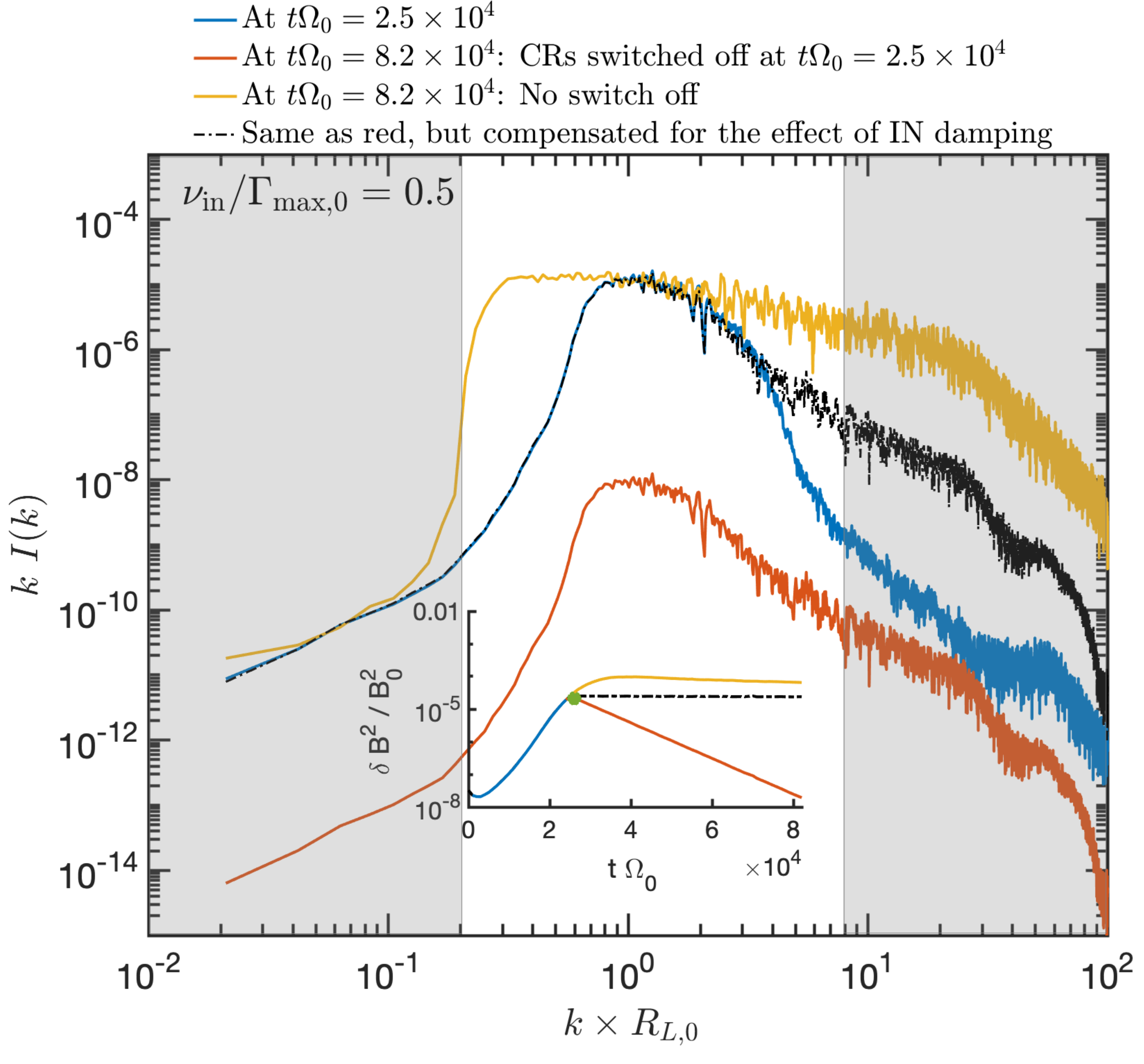}
\caption{Same as \autoref{fig:spectrum_swith_off} but with $\nu_{\rm in} / \Gamma_{\rm max, 0} = 0.5$. The additional dot-dashed black lines compensate analytically for the effect of ion-neutral damping on the wave spectrum. The gray-shaded areas delimit the regions where wave growth is not allowed in linear theory. (\autoref{eq:growth_rate_total}).} \label{fig:spectrum_swith_off_nuIN}
\end{center}
\end{figure}

The same effect is present in simulations with ion-neutral damping.  \autoref{fig:spectrum_swith_off_nuIN} is the same as \autoref{fig:spectrum_swith_off}  but with $\nu_{\rm in} / \Gamma_{\rm max, 0} = 0.5$. The additional dot-dashed black lines analytically compensate for the effect of ion-neutral damping on the wave spectrum. During the post-linear phase, a decrease of wave energy at any $k$ is imposed by the ion-neutral damping (difference between blue and red lines). If one compensates the effect of damping, we observe the same excess of high-$k$ modes at the end of the post-linear phase (dot-dashed black line) as for the undamped case.

Because these high-$k$ modes  can grow even in the absence of CRSI, we conclude that the driving mechanism is purely an MHD effect (see further discussion in Section \ref{sect:discussion}).

 \subsection{Saturated phase}
  \label{subsect:non_linear}
  
At the end of the post-linear phase exponential growth of the instability at all wavelengths is completed. By this time the CRs also fully experience the back-reaction from interacting with the waves they generated. In this section, we characterize the particle and wave properties in based on analysis of our MHD-PIC simulations at late stages. 

\begin{figure}
\begin{center}
\includegraphics[width=0.49\textwidth]{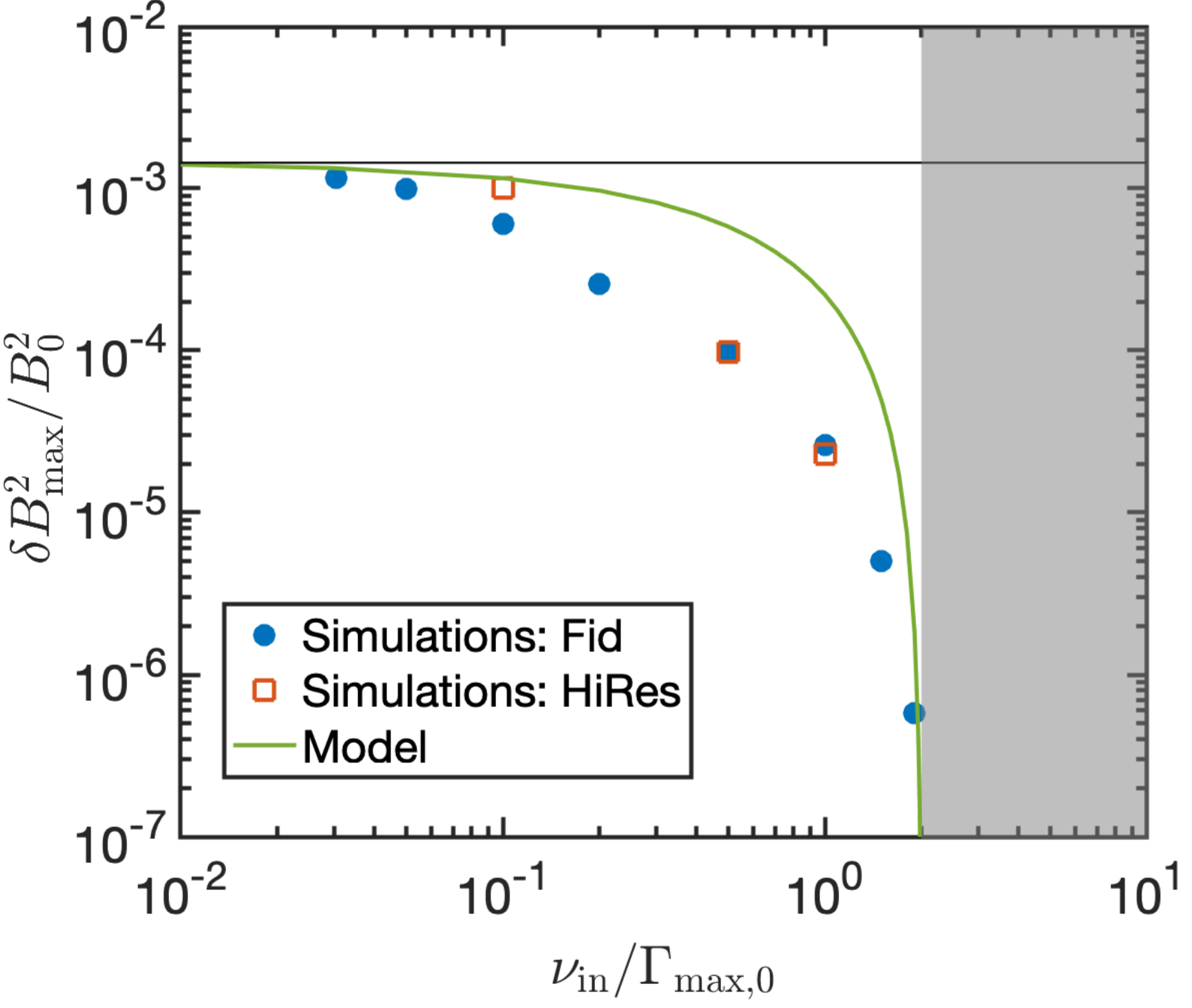}
\end{center}
\caption{Dependence of the maximum value of the magnetic wave energy, $\delta B^2_{\rm max} / B_0^2$, as function of $\nu_{\rm in}$. Blue circles are from Fid simulations and red squares from HiRes simulations. The solid green curve corresponds to the model solving two coupled ODEs, described in the text. The horizontal black line delimits the maximal allowed value at $\nu_{\rm in} = 0$, based on the measured saturation amplitude.  For $\nu_{\rm in} / \Gamma_{\rm max, 0} > 2$ there is no linear instability possible; this limit is shown by the gray-shaded region on the right side of the figure.} \label{fig:dB_sat_nuIN}
\end{figure}

\subsubsection{Late-stage wave amplitudes and particle distributions}

In \autoref{fig:dB_sat_nuIN} we present the maximum wave intensity (normalized to  $B_0^2$) reached in the simulations as function of $\nu_{\rm in}$ (plotted using blue circles and red squares for Fid and HiRes simulations, respectively). For the undamped or weakly damped cases we expect the momentum flux associated with the original anisotropy in the  
CRs to be transferred to forward-propagating Alfv\'en waves \citep[see][]{kulsrud_05}. As found in previous  simulations   
\citep[e.g., ][]{bai_19, holcomb_spitkovsky_18},
this leads to magnetic wave energy at saturation given by
$\delta B_{\rm sat}^2 / B_0^2 \approx 1.5 (n_{\rm CR} / n_i) \left(V_D/V_{A,i}-1\right)$.
This value is plotted as the horizontal black line in \autoref{fig:dB_sat_nuIN}.  The gray-shaded region at  $\nu_{\rm in} > 2 \Gamma_{\rm max, 0}$ delimits the region where the linear instability becomes impossible, according to \autoref{eq:growth_rate_total}.\footnote{We verified that there is no wave growth for $\nu_{\rm in} > 2 \Gamma_{\rm max, 0}$ with a dedicated simulation, not presented here.} The simulations show a gradual decrease in $(\delta B_{\rm max}/B_0)^2$ with increasing $\nu_{\rm in}$, as might be expected. The decrease becomes abrupt when approaching $\nu_{\rm in} = 2 \Gamma_{\rm max, 0}$, resembling an exponential cutoff.

In order to obtain some insight into the dependence of saturated wave intensity on $\nu_{\rm in}$, we cast our knowledge of dominant dynamical processes into a simple model. This model consists of a system of two coupled ODEs. Let $\mathcal{A}=(\delta B/B_0)^2$ be the wave amplitude squared, and $\Gamma$ be the wave growth rate (without damping). In our simple model, evolution of $A$ in time is determined by
\begin{eqnarray}
 \frac{ {\rm d} \left( \ln \mathcal{A} \right) }{{\rm d} t}    &=& 2 \left(\Gamma-\frac{\nu_{\rm in}}{2} \right) \, , \\
  \frac{ {\rm d} (\ln \Gamma)}{{\rm d} t}  &=& -\nu_{\rm s} \simeq - \mathcal{A} \Omega \frac{\pi}{8} g \, ,
\end{eqnarray}
 where $g$ is some factor of order unity. The first equation describes linear wave growth partially limited by IN damping. The second equation serves as a proxy for the dynamical adjustment of the growth rate to the changing CR distribution, which is becoming more isotropic under the effect of QLD in the bath of (growing) waves. For initial conditions, we set $\mathcal{A}(0)=10^{-10}$, $\Gamma(0)=\Gamma_{\rm max, 0}$.
We further parameterize IN damping rate as $\nu_{\rm in}=\alpha\Gamma_{\rm max, 0}$.
In our simulations, we have $\Gamma_{\rm max, 0} = 2.5\times 10^{-4} \Omega_0$, and $\alpha$
ranging from 0 to 2. We integrate these equations until $\Gamma/\Gamma_{\rm max, 0} < \alpha/2$
so that $\mathcal{A}$ reaches a maximum. In the case without damping, the wave amplitude should reach the value measured in simulations; from this constraint we find $g \approx 1.2$.

The solutions for $\mathcal{A}$ as function of $\nu_{\rm in}$ are plotted with a green solid line in \autoref{fig:dB_sat_nuIN}.
The model qualitatively agrees with the simulation results and shows a good match with our high-resolution simulation results when damping is weak ($\nu_{\rm in} \lesssim 0.1\Gamma_{\rm max,0}$), but  overpredicts $\delta B_{\rm max}$ in the case of moderate-to-strong damping ($ 0.2 <\nu_{\rm in} / \Gamma_{\rm max, 0} < 1$).

While a better match can be achieved with additional parameters and fine-tuning, this does not necessarily add further insight into the physical processes involved. In particular, we note that an important effect not captured in our toy model is the $\mu=0$ barrier (which is partly physical and partly numerical). When the isotropization process becomes stuck as particles accumulate at the barrier, further wave growth is suppressed, even though the free energy from the CR anisotropy has not been fully utilized. 

Indeed, in  \autoref{fig:Fdist_resolution_dependence}, it is evident that the HiRes model with $\nu_{\rm in}/\Gamma_{\rm max, 0} = 0.1$ (which shows good agreement with the toy model in 
\autoref{fig:dB_sat_nuIN}) 
does not have appreciable particle buildup at $\mu=0$, but the $\nu_{\rm in}/\Gamma_{\rm max, 0} = 0.5$ model (which falls below the model prediction) does have particles built up at $\mu=0$.

\begin{figure*}
\begin{center}
\includegraphics[width=0.97\textwidth]{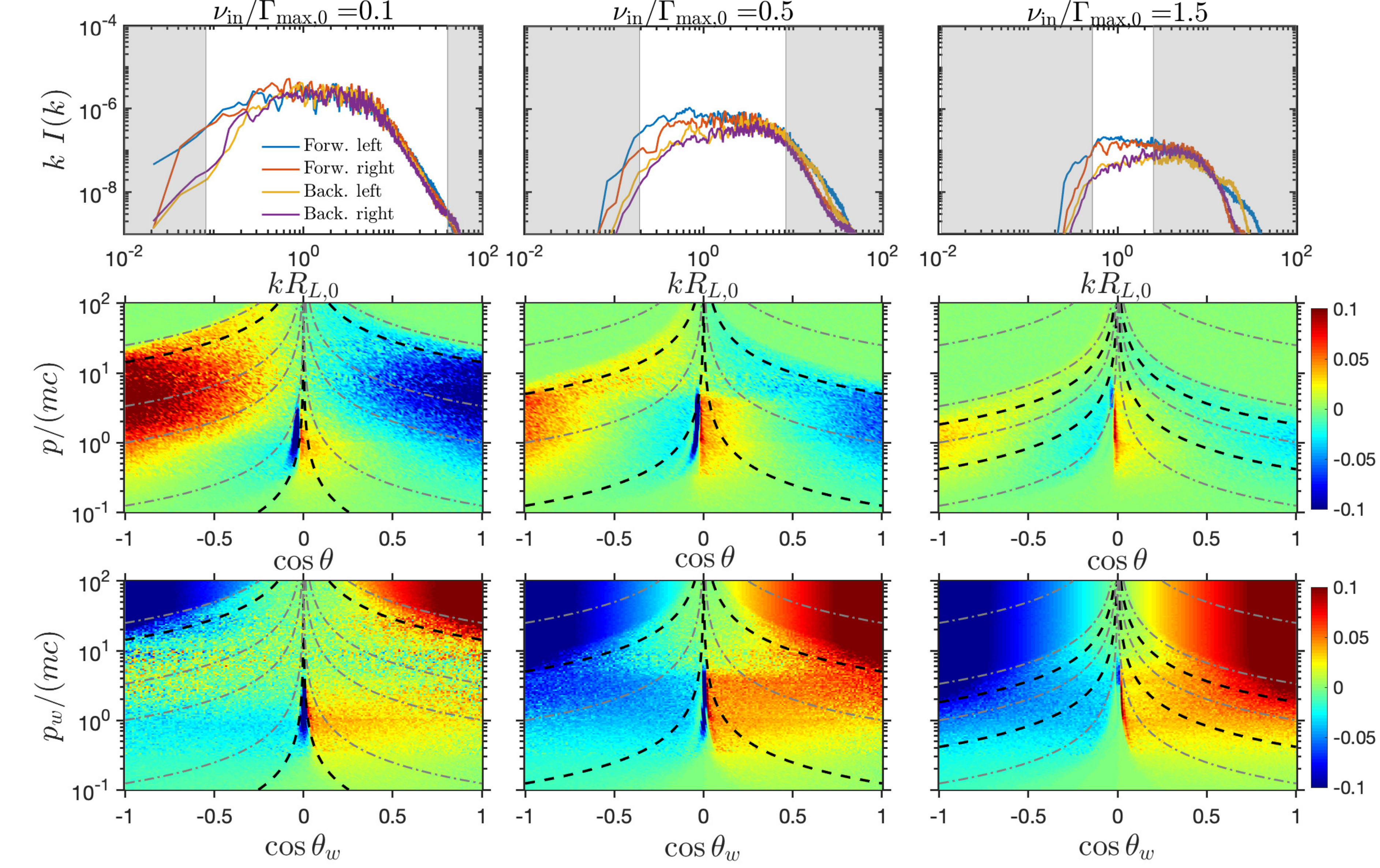}
\end{center}
\caption{Final spectrum (top panels), distribution function in the simulation frame $\delta f(p, \cos \theta)/f_0$ (middle panels), and  distribution function in the wave frame  $\delta f_w(p_w, \cos \theta_w)/f_0$ (bottom panels) in three simulations with different $\nu_{\rm in}$. Left, middle, and right columns have $\nu_{\rm in}/\Gamma_{\rm max, 0} = 0.1$, $0.5$ and $1.5$, respectively. The gray shaded regions in the upper panels delimit the regions where waves cannot grow during the linear phase, according to \autoref{eq:growth_rate_total}. Thick dashed lines in middle and bottom panels are the contours in momentum space resonant with the limiting  $k$ (i.e. the inner border of gray-shaded areas in the upper panels). The gray dot-dashed lines illustrate resonant contours for a few different wavenumbers.
} \label{fig:multipanel_final_nuINs}
\end{figure*}

\autoref{fig:multipanel_final_nuINs} presents the final state (at  $t\Omega_0 = 10^6$) of three representative simulations with different values of $\nu_{\rm in}$:  columns from left to right show models using $\nu_{\rm in}/\Gamma_{\rm max, 0} = 0.1$, $0.5$ and $1.5$, respectively.  The gray-shaded regions in the top panels mark the regions where waves cannot grow during the linear phase, according to \autoref{eq:growth_rate_total}. 

Some interesting features of the saturated phase are apparent in \autoref{fig:multipanel_final_nuINs}. Firstly, there is a reduction in the global wave intensity with increasing IN damping rate (compare from left to right top panels). Secondly, the saturated wave spectrum width exceeds the range imposed by the linear growth, with significant wave amplitude in gray-shaded regions (best evidenced in top right panel). Also, the forward-propagating waves (blue and red lines in top panels) are at a similar level to the backward-propagating waves (yellow and purple lines in top panels), a marked reduction from peak values reached during the post-linear phase brings all modes to comparable intensity level.  High-momentum particles (regions with $p>10 p_0$ in middle and bottom panels) clearly isotropize less efficiently with increasing $\nu_{\rm in}$:  from left to right in bottom panels, there is an increasing area that is unaffected by the instability. This corresponds to the lack of waves in $k R_{L,0} \ll 1$ region that becomes more and more prominent when $\nu_{\rm in}$ approaches the critical value $2\Gamma_{\rm max, 0}$.  Finally, we observe an accumulation of particles near $\mu = 0$ in all cases with IN damping (
``hot spots'' in the region close to $\cos \theta = 0$ in the middle panels and close to $\cos \theta_w = 0$ in the bottom panels). The issue of crossing $\mu = \cos \theta =0$ barrier becomes crucial when non-negligible damping is present. Not only is the global level of waves reduced, but also the whole spectrum becomes narrower when $\nu_{\rm in}$ approaches $2 \Gamma_{\rm max, 0}$. Both effects contribute in making it difficult to scatter across $\mu=0$.

 \subsection{Dependence on spatial resolution}

By testing different grid resolutions we confirm that there is no noticeable effect on the linear phase of the instability: no difference in the growth rate, wave energy, streaming velocity, nor spectrum during this phase. However, the resolution can have an important impact on the post-linear phase. Specifically, resolving a larger dynamic range of wave modes with $k R_{L,0} \gg 1$ allows us to better capture the fast post-linear growth of these modes.   This can be seen by comparing the high-$k$ part of the 
power spectra at $t \Omega_0 =  10^5$ in Figure~\ref{fig:spectra_twocase_lin_to_nonlin} for the fiducial resolution to those at similar times ($t \Omega_0 = 8.2\times 10^4$, with CRs kept on) in \autoref{fig:spectrum_swith_off} and \autoref{fig:spectrum_swith_off_nuIN} at high resolution. 

\begin{figure}
\begin{center}
\includegraphics[width=0.45\textwidth]{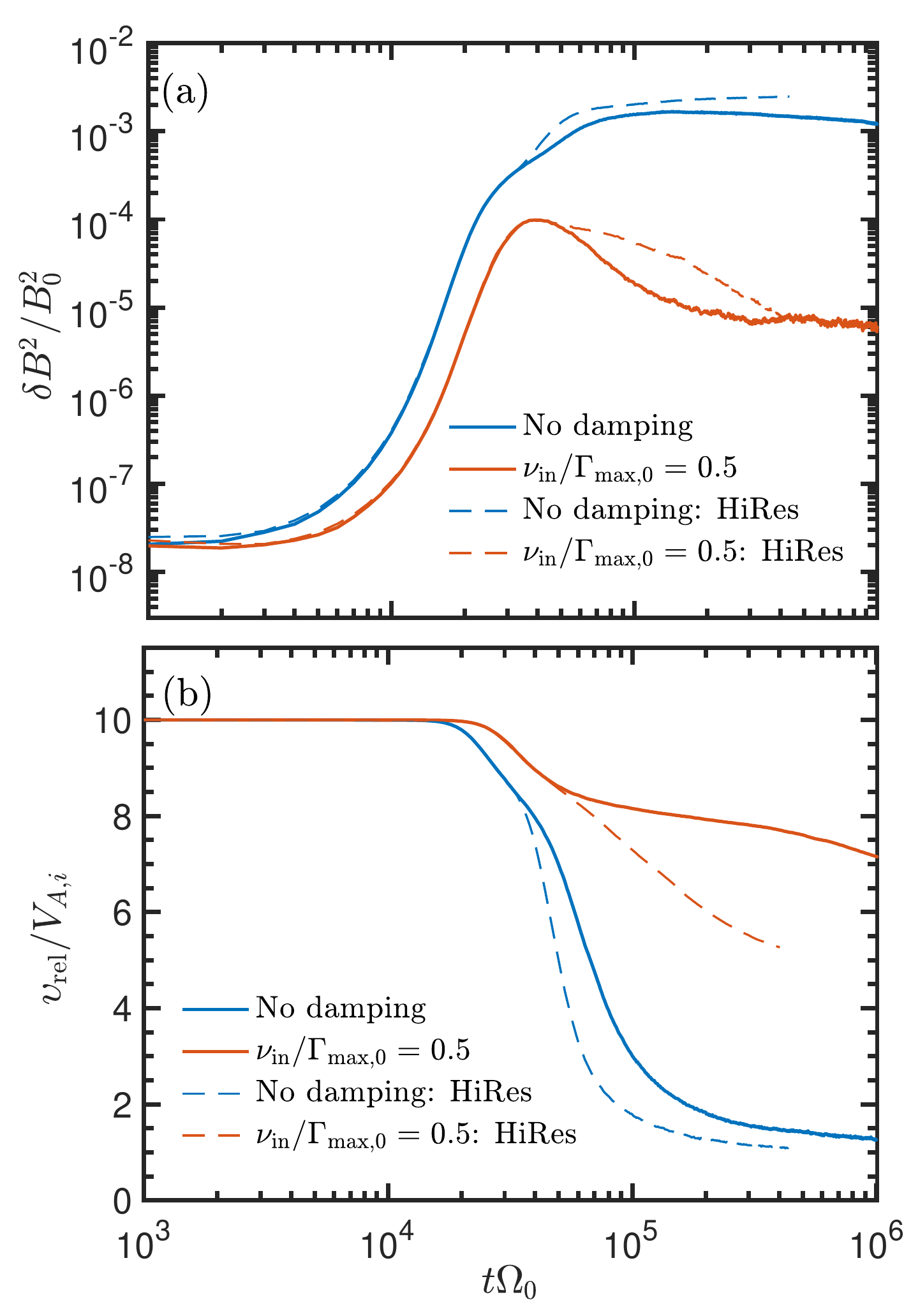}
\end{center}
\caption{Effect of numerical grid resolution on the time-evolution of the magnetic wave intensity (top panel) and of the CR streaming velocity (bottom panel). Blue lines show the case with no damping, and red lines show the case with $\nu_{\rm in} / \Gamma_{\rm max, 0} = 0.5$. Solid lines correspond to fiducial resolution and dashed lines correspond to high-resolution simulations (see \autoref{tab:params}).} \label{fig:dB_Vrel_resolution_dependence}
\end{figure}

In \autoref{fig:dB_Vrel_resolution_dependence} we compare the Fid and HiRes simulations for the time-evolution of the magnetic wave energy (top panel) and of the CR streaming velocity  (bottom panel). Two representative cases are shown: no damping (blue lines) and moderate damping $\nu_{\rm in} / \Gamma_{\rm max, 0} = 0.5$ (red lines). 
There is no difference in early evolution, until the post-linear phase, starting at $t \Omega_0 \sim 3-4\times 10^4$.  From the solid (Fid) and dashed (HiRes) blue lines, increased resolution has only a minor effect on the case with no damping. More noticeable differences appear for the case with moderate damping. During the post-linear phase, the magnetic energy decreases more slowly in HiRes simulation and the streaming velocity decreases at a faster rate. This effect is due to the higher level of wave intensity stored in high-$k$ modes during the post-linear phase.

\begin{figure}
\begin{center}
\includegraphics[width=0.45\textwidth]{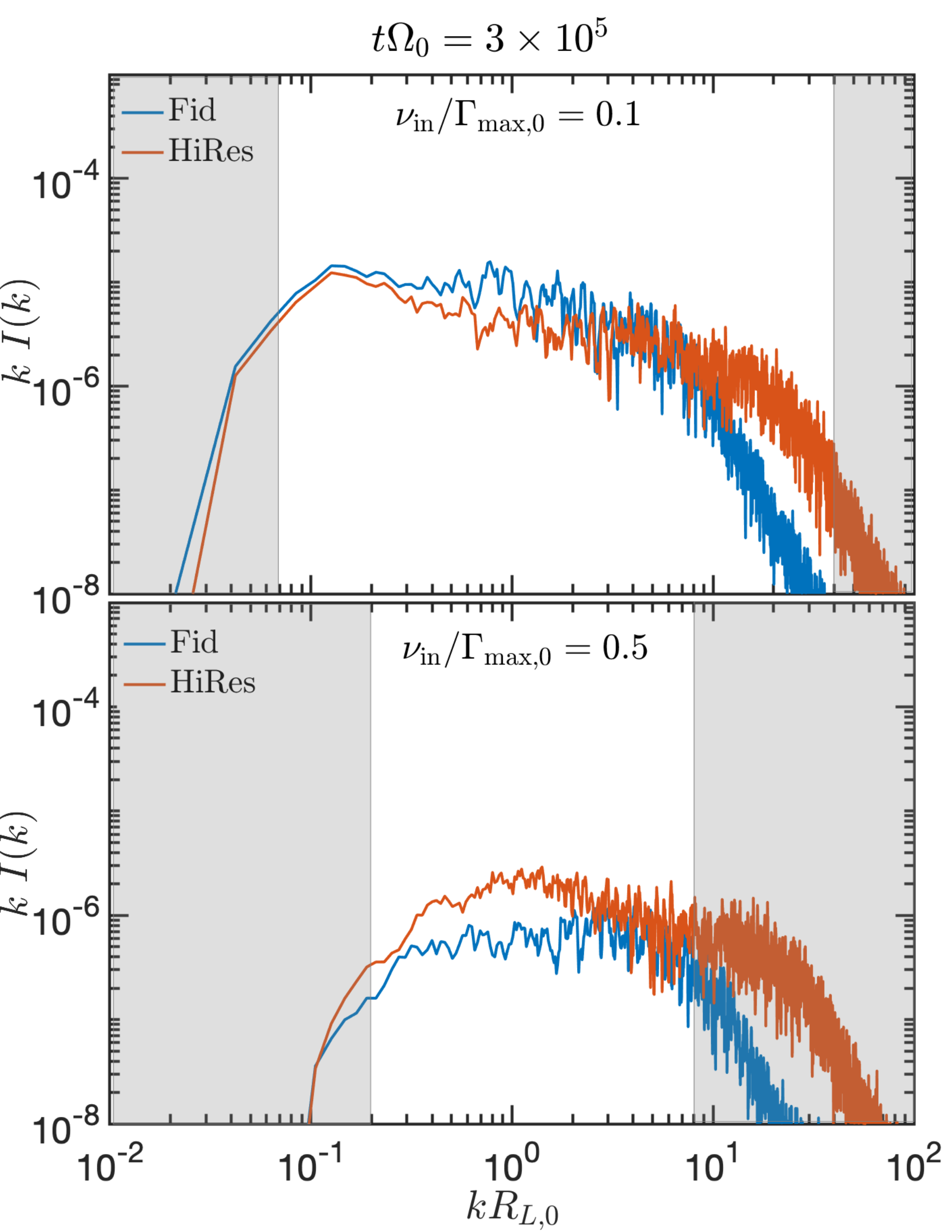}
\end{center}
\caption{Dependence on spatial resolution of the saturated wave spectrum (forward right-handed modes) for models with $\nu_{\rm in}/\Gamma_{\rm max, 0} = 0.1$ (top panel) and $\nu_{\rm in}/\Gamma_{\rm max, 0} = 0.5$ (bottom panel). Blue lines correspond to Fid simulations and red lines correspond to HiRes simulations.} \label{fig:spectrum_resolution_dependence}
\end{figure}

As found for the post-linear stage, numerical resolution has also an effect on the saturated state of the simulations. We expect, and indeed find,  that higher resolution simulations achieve a higher level of small-scale waves.  In \autoref{fig:spectrum_resolution_dependence} we present the comparison of the wave spectrum at $t\Omega_0 = 3 \times 10^5$ between Fid and HiRes simulations for two values of $\nu_{\rm in} = 0.1$ (top) and $0.5$ (bottom).
For both values of  $\nu_{\rm in}$ there are some difference around $k R_{L,0} = 1$. But, more importantly, the HiRes simulations exhibit higher level of wave intensity at $k R_{L,0} \gg 1$, without regard to the limit of linear growth (which is equal to 0 in gray-shaded areas).

\begin{figure}
\begin{center}
\includegraphics[width=0.45\textwidth]{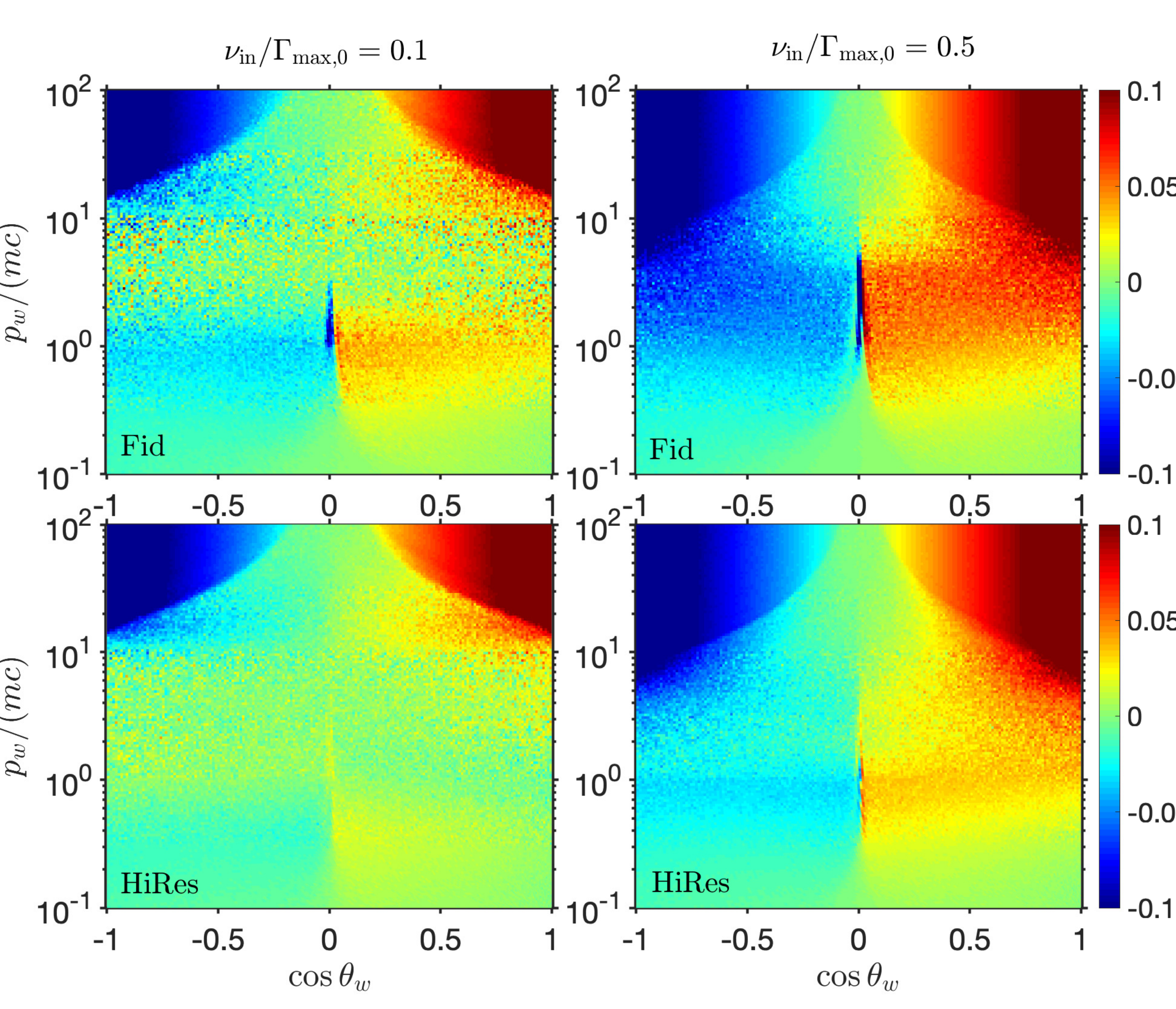}
\end{center}
\caption{Dependence on spatial resolution of the distribution function $\delta f_w (p_w, \cos \theta_w) / f_0$, at $t\Omega_0 = 3 \times 10^5$. Top panel corresponds to Fid simulations and bottom panel corresponds to HiRes simulations. Left and right show weak and moderate damping cases.} \label{fig:Fdist_resolution_dependence}
\end{figure}

The differences in the wave spectrum between Fid and HiRes simulations also lead to differences in the  final state of the CR distribution function.
In particular, higher amplitudes at large $k$ increase the scattering rate at the resonances close to $\mu = 0$, considering the resonance condition $k = (1/\mu)m_p\Omega/p$.
\autoref{fig:Fdist_resolution_dependence} presents the distribution function at  $t\Omega_0 = 3 \times 10^5$ of the simulations for Fid runs (upper panels) and HiRes runs (lower panels). We show cases with both weak (left) and moderate (right) damping, as in 
\autoref{fig:spectrum_resolution_dependence}.
For both, HiRes simulations show a better level of isotropization of $\delta f_w$ at $p/p_0 \in [0.5, 10]$, corresponding to the higher level of high-$k$ modes in HiRes simulations. No noticeable difference is observed at $p >10 p_0$, corresponding to quite similar spectra at $k R_{L,0} <0.1$ (see \autoref{fig:spectrum_resolution_dependence}).

\begin{figure}
\begin{center}
\includegraphics[width=0.45\textwidth]{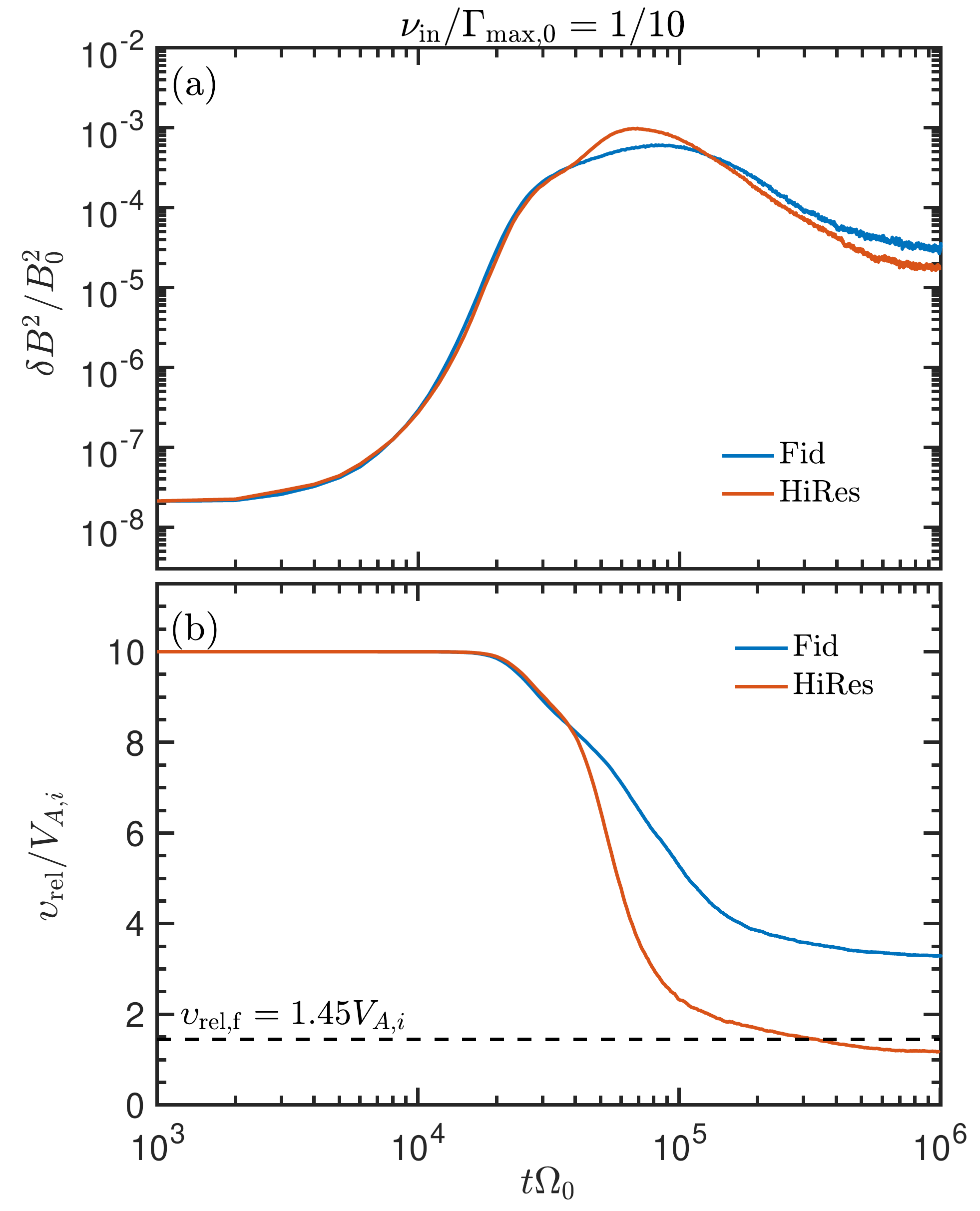}
\end{center}
\caption{Dependence on spatial resolution of the time-evolving wave energy (panel a) and streaming velocity (panel b), for $\nu_{\rm in}/\Gamma_{\rm max, 0}=1/10$. Blue and red line correspond to Fid and HiRes simulation, respectively. The horizontal dashed line delimits the final streaming velocity as expected by balancing the fastest growth rate with damping rate: $\Gamma_{\rm max, 0} = \nu_{\rm in}/2$.} 
\label{fig:Vrel_resolution_dependence_nu01}
\end{figure}

\subsection{Streaming velocity}

We now consider evolution of the CR bulk or streaming velocity, a key parameter since 
this characterizes the net CR flux. As mentioned in Section~\ref{subsec:Saturation_KC_criterion_and_QL_scattering},  in  the astrophysical literature the most commonly-adopted assumption is that the wave amplitude and streaming velocity are consistent with a state in which the linear growth rate balances the damping rate \citep{kulsrud_cesarsky_71}.  From Equations~\ref{eq:growth_rate_max} and \ref{eq:growth_rate_total}, this would imply that if the CRs remain isotropic in a frame moving at 
$V_{\rm st}$  relative to the gas, this velocity will decline until \autoref{eq:Streaming_velocity_Gmax} is satisfied. 

Conceptually, the asymptotic streaming velocity obtained through balancing damping and growth  relates to the situation where the instability is driven, i.e., through a CR pressure gradient. In contrast, our numerical setup with periodic boundary conditions corresponds more to a `transient’ situation in which instability dies out.  This affects the astrophysical relevance of the streaming speed derived in our simulations. Nevertheless, it is interesting to follow the measured time-evolution of CR streaming from the simulations.

In the bottom panel of \autoref{fig:Vrel_resolution_dependence_nu01} we show the evolution of measured CR streaming velocity $v_{\rm rel}$ (defined in \autoref{eq:streaming_velocity_def}), including its dependence on numerical resolution, for the weak damping case $\nu_{\rm in}/\Gamma_{\rm max, 0} = 0.1$. In the Fid simulation (blue line) the final streaming velocity of CRs is $3.3 V_{A,i}$. The HiRes simulation isotropizes more efficiently, showing a decrease of the streaming velocity to $v_{\rm rel} = 1.18 V_{A,i}$  (but still above full isotropization $v_{\rm rel} = V_{A,i}$).  We attribute this difference to better capturing of high-$k$ modes in HiRes simulations.

If we were to apply \autoref{eq:Streaming_velocity_Gmax} to the case with $\nu_{\rm in} =0.1 \Gamma_{\rm max, 0}$, the predicted asymptotic streaming velocity would be $V_{\rm st}/V_{\rm A,i} = 1.45$.  This level is marked by the dashed horizontal line in \autoref{fig:Vrel_resolution_dependence_nu01}. Interestingly, for the HiRes simulation the measured final-time  $v_{\rm rel}$ is below this. This can be understood from \autoref{fig:spectrum_resolution_dependence}: as long as a sufficient level of waves is present the distribution will continue to isotropize and $v_{\rm rel}$ will continue to decline.  In constrast, for the HiRes model with $\nu_{\rm in} =0.5\Gamma_{\rm max, 0}$  (see Figure~\ref{fig:dB_Vrel_resolution_dependence}, red lines), the late-time $v_{\rm rel}$ remains well above the value $3.25 V_{A,i}$ that would be predicted by \autoref{eq:Streaming_velocity_Gmax},   presumably due to the lower wave amplitudes that reduce the scattering rate (compare upper and lower panels of Figure~\ref{fig:spectrum_resolution_dependence}, and note also the buildup of particles to the right of $\mu=0$ in the lower-right of Figure~\ref{fig:Fdist_resolution_dependence}).
Taken together, these results make clear that the simple approach of  setting $\Gamma_{\rm max, 0} =\nu_{\rm in}/2$ 
and solving for streaming velocity is not necessarily applicable. The evolution and final value of  $v_{\rm rel}$ instead depends on the evolution of $(\delta B/B_0)^2$.  This is illustrated for different values of $\nu_{\rm in}$ in \autoref{fig:dB_Vrel_resolution_dependence}, with more complete isotropization in cases of lower $\nu_{\rm in}$.  In the real astrophysical case, an additional factor affecting evolution would be the history of (anisotropic) energy sources.

Finally, we recall (as pointed out in Section~\ref{subsec:Saturation_KC_criterion_and_QL_scattering}) that balancing the linear growth time and damping time of waves does not quantitatively take into account the additional time required for QLD to isotropize the distribution, suggesting that the criterion $\Gamma_{\rm max, 0} >\nu_{\rm in}/2$ is necessary but not sufficient. Because the pitch angle diffusion timescale
near saturation is approximately proportional
to the linear growth timescale, there may still be some critical ratio of growth to damping that allows the distribution to isotropize.  This question could be addressed  in future work that allows for more realistic driving. 

%%%%%%%%%%%%%%%%%%%%%%%%%%%%%%%%%%%%%%%%%%%%%%%%%%%%%%%%%%%%%%%%%%
% New section
%%%%%%%%%%%%%%%%%%%%%%%%%%%%%%%%%%%%%%%%%%%%%%%%%%%%%%%%%%%%%%%%%%
  
\section{Discussion}
\label{sect:discussion}

Perhaps one of the most intriguing results of the present work is the evidence of high-$k$ growth during the post-linear phase, presented in section \ref{subsect:post_linear}. In Figures \ref{fig:spectrum_swith_off} and \ref{fig:spectrum_swith_off_nuIN} we have shown that these modes do not require driving by CRs to be amplified. Presently, we do not have a satisfactory explanation of this effect, but hypothesize that it is due to some form of mode coupling or wave steepening into rotational discontinuities \citep[e.g.,][]{cohen_74}. One could argue that this effect is due to our choice of parameters that lead to relatively high wave amplitude at saturation, up to  $\delta B / B_0 \simeq 0.05$. This is likely considerably  larger than is present in the general ISM. However, we also performed simulations with $n_{\rm CR} / n_i$ as small as $5\times 10^{-5}$ and $V_D / V_{A,i} = 2$, observing the same effect of fast high-$k$ mode growth during the post-linear phase. Potentially, the non-linearity or three-wave interactions could be diagnosed by bispectrum and bicoherence analysis, 
but we defer this exploration to future work.

We also found that large-wavelength modes with $k R_{L,0} \ll 1$ are not amplified if CR driving is switched off at the end of the linear phase of the instability. This shows that these modes are mainly amplified by resonant interaction with CRs, with growth rate $\Gamma(k) \propto k^{2\kappa-1}$, and that the IN damping removes the low-$k$ end of the wave spectrum  with $k < k_{\rm min}$, where $\Gamma(k < k_{\rm min}) < \nu_{\rm in} / 2$.  This can be seen in Figures \ref{fig:spectra_twocase_lin_to_nonlin}, \ref{fig:spectrum_swith_off_nuIN}, \ref{fig:multipanel_final_nuINs}. This implies, assuming only resonant wave-particle interaction, that high-momentum CRs with $p > p_{\rm max}$ are not isotropized and continue to freely stream. Here, $p_{\max}$ is deduced by using \autoref{eq:growth_rate_total} with \autoref{eq:grate_kappa} using the resonance condition $k R_{L,0} = p_0 /(p\mu)$: 

\be
\frac{(p\mu)_{\rm max}}{p_0} \simeq \left[ 1.55  \frac{n_{\rm CR}}{n_i} \frac{\Omega_i}{\nu_{\rm in}} \left( \frac{V_D}{V_{A,i}} - 1\right) \right]^{\frac{2}{3}} \, ,
\label{eq:Emax_for_isotropy}
\ee
where we fixed $\kappa = 1.25$.

Similarly to \citet{kulsrud_cesarsky_71} we estimate the maximum momentum of CRs that could potentialaly be isotropized (i.e., self-confined) in different phases of the ISM. Using typical values given in \autoref{tab:ISM_phases} and \autoref{eq:Emax_for_isotropy}, the derived values of $(p\mu)_{\rm max}$ are given in \autoref{tab:Pmax_phases}, adopting $V_D/V_{A,i} = 2$ and $V_D = 0.1 c$ as two extreme cases. Assuming $p_0 =m_p c$ for CRs with energy $E \simeq GeV$, and we draw some general conclusions:
\begin{itemize}
\item \textbf{DMG}: strongly suppressed CRSI; no CR isotropization at energies larger than GeV.  Streaming could become extremely large. 
\item \textbf{MG}: marginal instability and isotropy at GeV but not beyond 100 GeV. The surface layers of molecular clouds may therefore be subject to  CRSI if the initial anisotropy of CRs is not too small.
\item \textbf{CNM}: the conditions for triggering the CRSI are only satisfied for highly super-Alfv\'enic streaming.  The CR distribution could become quite anisotropic in this phase.
\item \textbf{WNM}: most favourable environment for CRSI. Even with tiny anisotropy  ($V_D/V_{A,i}$ of order unity) the instability is possible for GeV CRs. CRs with $E\gg$ GeV could also be isotropized if the initial drift velocity is much larger than $V_{A,i}$.
\end{itemize}

In our own Milky Way and similar galaxies, the WNM is the dominant component of the ISM by mass; the above estimate affirms that CRSI is astrophysically quite important.  We note, however, that in the WNM other wave damping mechanisms can compete with IN damping that could prevent isotropization of CRs with energies $\geq$TeV \citep[see, e.g., Fig.1 and corresponding text in][]{brahimi_20}, but our values of $E_{\rm max}$ are in good agreement with similar calculation by \citet{xu_2016} (their Sect. 6.4 and Fig.15). 

The estimates above are in broad agreement with those of \citet{kulsrud_cesarsky_71} -- being based on the same argument -- but are updated here with representative parameters for different neutral-dominated media. 

\begin{table}
\caption{Maximum momentum of cosmic-rays able to self-confine for different phases of the ISM.}\label{tab:Pmax_phases}
\begin{center}
\begin{tabular}{c|cc}\hline\hline
 Phase & $\frac{(p\mu)_{\rm max}}{p_0}$, $V_D=2V_{A,i}$ & $\frac{(p\mu)_{\rm max}}{p_0}$, $V_D=0.1 c$\\ \hline
WNM & 12.2& 345  \\
CNM & 0.34 & 22.1 \\
MG & 1.5 & 41.7  \\
DMG & $\ll 1$ & $<1$ \\
\hline\hline
\end{tabular}
\end{center}
Note -- Other parameter values are as in \autoref{tab:ISM_phases}.
\end{table}

The high-frequency limit was adopted in the present study, i.e., $\omega_A = k V_{A,i} \gg \max [ \nu_{\rm in}, \nu_{\rm ni}]$. In this regime, the neutral fluid and ionized fluid are decoupled, and the ion-neutral wave damping rate does not depend on the wavelength, $\Gamma_{\rm d} = \nu_{\rm in} / 2$. We expect this approximation to hold for any $k R_{L,0} \geq 0.1 - 1$, in general, and to be satisfied for CRs at GeV energies and below, which make up most of the CR  energy density and are responsible for most of the ionization. This justifies the one-fluid approach, adopted in the present study. However, in reality very  high energy CRs with low resonant frequencies are present in the ISM as well. If the whole CR  energy spectrum were fully  represented, the low-$k$ part of the wave spectrum (modes with $k R_{L,0} \ll max[\nu_{\rm in}, \nu_{\rm ni}]$) would be damped at slower rate than $\nu_{\rm in} / 2$. This could have an interesting effect on the amplification and survival of long-wavelength modes. However, we defer investigation of $k$-dependent damping to future study.

Similar to \citetalias{bai_19}, many of our simulations show particle accumulation at $\mu=0$.  Naively, this would be expected if waves were only present where CRSI is undamped (at intermediate $k$ near $k_{\rm max}= \sqrt{2 -1/\kappa} R_{L,0}^{-1}$), since there would be no large-$k$ waves that are able to scatter small-$\mu$ particles subject to the resonance condition $k \mu  = \Omega m/p$.  However, in practice nonlinear MHD effects populate the large-$k$ region regardless of whether CRSI is damped or active, and therefore the $\mu=0$ crossing is possible.  Nevertheless, in simulations the high-$k$ regime is subject to numerical dissipation, and we find that higher than standard resolution is required to limit particle buildup.  At  the higher values of $\nu_{\rm in}$,  particles still build up near $\mu=0$ even with higher resolution (see Figure~\ref{fig:Fdist_resolution_dependence}).

As we have previously emphasized, the problem of relating  the streaming rate and wave amplitude to macroscopic ambient properties  clearly merits further study. The traditional ``detailed balance'' approach of equating the CRSI growth rate to the damping rate has recently been adopted in galaxy formation simulations and other studies as a procedure for setting the scattering rate coefficient, leading to a diffusion coefficient that varies proportional to $\nu_{\rm in}$ \citep[see, e.g][]{hopkins_21b}. To assess and quantitatively improve this kind of presciption, however, studies similar to the present one but allowing for macroscopic driving (via an imposed CR  flux or energy gradient) will be needed.

%%%%%%%%%%%%%%%%%%%%%%%%%%%%%%%%%%%%%%%%%%%%%%%%%%%%%%%%%%%%%%%%%%
% New section
%%%%%%%%%%%%%%%%%%%%%%%%%%%%%%%%%%%%%%%%%%%%%%%%%%%%%%%%%%%%%%%%%%

\section{Conclusion}
\label{sect:conclusion}

Motivated by the recent progress in coupled fluid-kinetic (MHD-PIC) numerical techniques \citepalias{bai_19}, in this work we studied the influence of ambipolar diffusion (ion-neutral damping of Alfv\'en waves) on the CRSI.  
We adopted parameters for a reference model  that in the absence ion-neutral damping lead to full isotropization of the CR distribution  function in the wave frame after the instability saturates. This state corresponds to an asymptoptic state in which the CRs stream at Alfv\'en speed relative to the background fluid. We then explored different values of ion-neutral damping rate to study the influence on the outcome of the CRSI. 

Our main conclusions are summarized below:

\begin{itemize}
    \item The predicted exponential growth rate of the instability including IN damping, $\Gamma(k)$, is well-reproduced by the MHD-PIC technique.  Thus the linear theory is in good agreement with analytical expectations.
    \item Evolution in a post-linear phase is  crucial for isotropization of low- and moderate-energy CRs. During this phase, growth of high-$k$ (wavelength $\lambda \ll R_{L,0}$) modes occurs. This growth is not driven by CR anisotropy (the waves are outside of the unstable range) but rather appears to be a result of an MHD wave cascade or wave steepening into rotational discontinuities.  The high-$k$ waves are important to scattering when $\mu$ is close to 0. The development of the high-$k$ spectrum  is best captured in high-resolution simulations.
    \item Systematic comparison between the reference case (no damping) and simulations with IN damping reveals that the width of the wave spectrum decreases with increasing $\nu_{\rm in}$. 
    The absence of low-$k$ (compared to $R_{L,0}^{-1}$) waves prevents isotropization of high-energy CRs. 
    There is no wave growth at all if $\nu_{\rm in} > 2 \Gamma_{\rm max, 0}$, consistent with analytic theory.  
    \item With stronger IN damping, the maximum amplitude of waves systematically decreases.  When $\nu_{\rm in}=0.5 \Gamma_{\rm max, 0}$,  the peak wave amplitude is an order of magnitude below the no-damping case. Lower wave amplitudes reduce the rate of QLD and hence slows isotropization.
\end{itemize}

 This work is the second study in a series exploring the physics of CRSI by means of MHD-PIC simulations.   
 This numerical technique has proved to be valuable because it captures the relevant microphysical CR-gyroscales and allows long-term simulations at low CR density more affordably than with full kinetic simulations. We envision that further improvements of the method and inclusion of additional physics will lead to greater understanding of the microphysics of CR-ISM interactions in different phases, also enabling numerical calibrations of transport coefficients required for large-scale ISM/galactic studies. 
\bigskip

\acknowledgements
We are grateful to the anonymous referee for a careful reading of our manuscript and insightful report.
IP was supported by NSF grants PHY-1804048, PHY-1523261 and by the Max-Planck/Princeton Center for Plasma Physics.
The work of ECO  was supported by grant 510940 from the Simons Foundation.
XNB acknowledges support by NSFC grant 11873033.
Computations were conducted on resources at PICSciE-OIT High Performance Computing Center and Visualization Laboratory at Princeton University, on CALMIP supercomputing resources at Universit\'e de Toulouse-III (France) under the allocations 2016-p1504 and P20028, and on the Orion supercomputer through Department of Astronomy at Tsinghua University.

\bibliographystyle{aasjournal}
\bibliography{CRSIIN_biblio}

\begin{thebibliography}{}
\expandafter\ifx\csname natexlab\endcsname\relax\def\natexlab#1{#1}\fi
\providecommand{\url}[1]{\href{#1}{#1}}

\bibitem[{{Amano}(2018)}]{amano_18}
{Amano}, T. 2018, Journal of Computational Physics, 366, 366

\bibitem[{{Amato} \& {Blasi}(2009)}]{amato_blasi_09}
{Amato}, E., \& {Blasi}, P. 2009, \mnras, 392, 1591

\bibitem[{{Amato} \& {Blasi}(2018)}]{2018AdSpR..62.2731A}
---. 2018, Advances in Space Research, 62, 2731

\bibitem[{{Bai} {et~al.}(2015){Bai}, {Caprioli}, {Sironi}, \&
  {Spitkovsky}}]{bai_15}
{Bai}, X.-N., {Caprioli}, D., {Sironi}, L., \& {Spitkovsky}, A. 2015, \apj,
  809, 55

\bibitem[{{Bai} {et~al.}(2019){Bai}, {Ostriker}, {Plotnikov}, \&
  {Stone}}]{bai_19}
{Bai}, X.-N., {Ostriker}, E.~C., {Plotnikov}, I., \& {Stone}, J.~M. 2019, \apj,
  876, 60 {\bf [Paper I]}

\bibitem[{{Bambic} {et~al.}(2021){Bambic}, {Bai}, \& {Ostriker}}]{bambic_21}
{Bambic}, C.~J., {Bai}, X.-N., \& {Ostriker}, E.~C. 2021, arXiv e-prints,
  arXiv:2102.11877

\bibitem[{{Bell}(2004)}]{bell_04}
{Bell}, A.~R. 2004, \mnras, 353, 550

\bibitem[{{Berezinskii} {et~al.}(1990){Berezinskii}, {Bulanov}, {Dogiel}, \&
  {Ptuskin}}]{berezinskii_90}
{Berezinskii}, V.~S., {Bulanov}, S.~V., {Dogiel}, V.~A., \& {Ptuskin}, V.~S.
  1990, {Astrophysics of cosmic rays}

\bibitem[{{Blasi} {et~al.}(2012{\natexlab{a}}){Blasi}, {Amato}, \&
  {Serpico}}]{blasi_2012a}
{Blasi}, P., {Amato}, E., \& {Serpico}, P.~D. 2012{\natexlab{a}}, \prl, 109,
  061101

\bibitem[{{Blasi} {et~al.}(2012{\natexlab{b}}){Blasi}, {Morlino}, {Bandiera},
  {Amato}, \& {Caprioli}}]{blasi_12}
{Blasi}, P., {Morlino}, G., {Bandiera}, R., {Amato}, E., \& {Caprioli}, D.
  2012{\natexlab{b}}, \apj, 755, 121

\bibitem[{{Brahimi} {et~al.}(2020){Brahimi}, {Marcowith}, \&
  {Ptuskin}}]{brahimi_20}
{Brahimi}, L., {Marcowith}, A., \& {Ptuskin}, V.~S. 2020, \aap, 633, A72

\bibitem[{{Breitschwerdt} {et~al.}(1991){Breitschwerdt}, {McKenzie}, \&
  {Voelk}}]{breitschwerdt_91}
{Breitschwerdt}, D., {McKenzie}, J.~F., \& {Voelk}, H.~J. 1991, \aap, 245, 79

\bibitem[{{Bustard} \& {Zweibel}(2020)}]{bustard_20}
{Bustard}, C., \& {Zweibel}, E.~G. 2020, arXiv e-prints, arXiv:2012.06585

\bibitem[{{Butsky} \& {Quinn}(2018)}]{butsky_18}
{Butsky}, I.~S., \& {Quinn}, T.~R. 2018, \apj, 868, 108

\bibitem[{{Bykov} {et~al.}(2013){Bykov}, {Brandenburg}, {Malkov}, \&
  {Osipov}}]{bykov_13}
{Bykov}, A.~M., {Brandenburg}, A., {Malkov}, M.~A., \& {Osipov}, S.~M. 2013,
  \ssr, 178, 201

\bibitem[{{Bykov} \& {Toptygin}(2005)}]{bykov_toptygin_05}
{Bykov}, A.~M., \& {Toptygin}, I.~N. 2005, Astronomy Letters, 31, 748

\bibitem[{{Cohen} \& {Kulsrud}(1974)}]{cohen_74}
{Cohen}, R.~H., \& {Kulsrud}, R.~M. 1974, Physics of Fluids, 17, 2215

\bibitem[{{Dashyan} \& {Dubois}(2020)}]{dashyan_20}
{Dashyan}, G., \& {Dubois}, Y. 2020, \aap, 638, A123

\bibitem[{{Draine}(2011)}]{draine_11}
{Draine}, B.~T. 2011, {Physics of the Interstellar and Intergalactic Medium}

\bibitem[{{Dubois} \& {Commer{\c{c}}on}(2016)}]{dubois_16}
{Dubois}, Y., \& {Commer{\c{c}}on}, B. 2016, \aap, 585, A138

\bibitem[{{Dubois} {et~al.}(2019){Dubois}, {Commer{\c{c}}on}, {Marcowith}, \&
  {Brahimi}}]{dubois_19}
{Dubois}, Y., {Commer{\c{c}}on}, B., {Marcowith}, A.~r., \& {Brahimi}, L. 2019,
  \aap, 631, A121

\bibitem[{{Everett} \& {Zweibel}(2011)}]{everett_zweibel_11}
{Everett}, J.~E., \& {Zweibel}, E.~G. 2011, \apj, 739, 60

\bibitem[{{Everett} {et~al.}(2008){Everett}, {Zweibel}, {Benjamin}, {McCammon},
  {Rocks}, \& {Gallagher}}]{everett_08}
{Everett}, J.~E., {Zweibel}, E.~G., {Benjamin}, R.~A., {et~al.} 2008, \apj,
  674, 258

\bibitem[{{Farmer} \& {Goldreich}(2004)}]{farmer_goldreich_04}
{Farmer}, A.~J., \& {Goldreich}, P. 2004, \apj, 604, 671

\bibitem[{{Ferri{\`e}re}(2001)}]{ferriere_01}
{Ferri{\`e}re}, K.~M. 2001, Reviews of Modern Physics, 73, 1031

\bibitem[{{Ginzburg} {et~al.}(1973){Ginzburg}, {Ptuskin}, \&
  {Tsytovich}}]{ginzburg_73}
{Ginzburg}, V.~L., {Ptuskin}, V.~S., \& {Tsytovich}, V.~N. 1973, \apss, 21, 13

\bibitem[{{Girichidis} {et~al.}(2018){Girichidis}, {Naab}, {Hanasz}, \&
  {Walch}}]{girichidis_18}
{Girichidis}, P., {Naab}, T., {Hanasz}, M., \& {Walch}, S. 2018, \mnras, 479,
  3042

\bibitem[{{Girichidis} {et~al.}(2016){Girichidis}, {Naab}, {Walch}, {Hanasz},
  {Mac Low}, {Ostriker}, {Gatto}, {Peters}, {W{\"u}nsch}, {Glover}, {Klessen},
  {Clark}, \& {Baczynski}}]{girichidis_16}
{Girichidis}, P., {Naab}, T., {Walch}, S., {et~al.} 2016, \apjl, 816, L19

\bibitem[{{Glassgold} {et~al.}(2012){Glassgold}, {Galli}, \&
  {Padovani}}]{2012ApJ...756..157G}
{Glassgold}, A.~E., {Galli}, D., \& {Padovani}, M. 2012, \apj, 756, 157

\bibitem[{{Grenier} {et~al.}(2015){Grenier}, {Black}, \&
  {Strong}}]{2015ARA&A..53..199G}
{Grenier}, I.~A., {Black}, J.~H., \& {Strong}, A.~W. 2015, \araa, 53, 199

\bibitem[{{Haggerty} {et~al.}(2019){Haggerty}, {Caprioli}, \&
  {Zweibel}}]{haggerty_19}
{Haggerty}, C., {Caprioli}, D., \& {Zweibel}, E. 2019, in International Cosmic
  Ray Conference, Vol.~36, 36th International Cosmic Ray Conference (ICRC2019),
  279

\bibitem[{{Hanasz} \& {Lesch}(2003)}]{hanasz_03}
{Hanasz}, M., \& {Lesch}, H. 2003, \aap, 412, 331

\bibitem[{{Hanasz} {et~al.}(2013){Hanasz}, {Lesch}, {Naab}, {Gawryszczak},
  {Kowalik}, \& {W{\'o}lta{\'n}ski}}]{hanasz_13}
{Hanasz}, M., {Lesch}, H., {Naab}, T., {et~al.} 2013, \apjl, 777, L38

\bibitem[{{Holcomb} \& {Spitkovsky}(2019)}]{holcomb_spitkovsky_18}
{Holcomb}, C., \& {Spitkovsky}, A. 2019, \apj, 882, 3

\bibitem[{{Hopkins} {et~al.}(2021{\natexlab{a}}){Hopkins}, {Chan}, {Squire},
  {Quataert}, {Ji}, {Kere{\v{s}}}, \& {Faucher-Gigu{\`e}re}}]{hopkins_21a}
{Hopkins}, P.~F., {Chan}, T.~K., {Squire}, J., {et~al.} 2021{\natexlab{a}},
  \mnras, 501, 3663

\bibitem[{{Hopkins} {et~al.}(2021{\natexlab{b}}){Hopkins}, {Squire}, {Chan},
  {Quataert}, {Ji}, {Kere{\v{s}}}, \& {Faucher-Gigu{\`e}re}}]{hopkins_21b}
{Hopkins}, P.~F., {Squire}, J., {Chan}, T.~K., {et~al.} 2021{\natexlab{b}},
  \mnras, 501, 4184

\bibitem[{{Hopkins} {et~al.}(2020){Hopkins}, {Chan}, {Garrison-Kimmel}, {Ji},
  {Su}, {Hummels}, {Kere{\v{s}}}, {Quataert}, \&
  {Faucher-Gigu{\`e}re}}]{hopkins_19}
{Hopkins}, P.~F., {Chan}, T.~K., {Garrison-Kimmel}, S., {et~al.} 2020, \mnras,
  492, 3465

\bibitem[{{Ipavich}(1975)}]{ipavich_75}
{Ipavich}, F.~M. 1975, \apj, 196, 107

\bibitem[{{Ivlev} {et~al.}(2018){Ivlev}, {Dogiel}, {Chernyshov}, {Caselli},
  {Ko}, \& {Cheng}}]{ivlev_18}
{Ivlev}, A.~V., {Dogiel}, V.~A., {Chernyshov}, D.~O., {et~al.} 2018, \apj, 855,
  23

\bibitem[{{Jiang} \& {Oh}(2018)}]{jiang_oh_18}
{Jiang}, Y.-F., \& {Oh}, S.~P. 2018, \apj, 854, 5

\bibitem[{{Krumholz} {et~al.}(2020){Krumholz}, {Crocker}, {Xu}, {Lazarian},
  {Rosevear}, \& {Bedwell-Wilson}}]{krumholz_20}
{Krumholz}, M.~R., {Crocker}, R.~M., {Xu}, S., {et~al.} 2020, \mnras, 493, 2817

\bibitem[{{Kulsrud} \& {Pearce}(1969)}]{kulsrud_pearce_69}
{Kulsrud}, R., \& {Pearce}, W.~P. 1969, \apj, 156, 445

\bibitem[{{Kulsrud}(1978)}]{kulsrud_78}
{Kulsrud}, R.~M. 1978, in Astronomical Papers Dedicated to Bengt Stromgren, ed.
  A.~{Reiz} \& T.~{Andersen}, 317--326

\bibitem[{{Kulsrud}(2005)}]{kulsrud_05}
{Kulsrud}, R.~M. 2005, {Plasma physics for astrophysics}

\bibitem[{{Kulsrud} \& {Cesarsky}(1971)}]{kulsrud_cesarsky_71}
{Kulsrud}, R.~M., \& {Cesarsky}, C.~J. 1971, \aplett, 8, 189

\bibitem[{{Lazarian}(2016)}]{lazarian_16}
{Lazarian}, A. 2016, \apj, 833, 131

\bibitem[{{Lebiga} {et~al.}(2018){Lebiga}, {Santos-Lima}, \& {Yan}}]{lebiga_18}
{Lebiga}, O., {Santos-Lima}, R., \& {Yan}, H. 2018, \mnras, 476, 2779

\bibitem[{{Lee} \& {V{\"o}lk}(1973)}]{lee_volk_73}
{Lee}, M.~A., \& {V{\"o}lk}, H.~J. 1973, \apss, 24, 31

\bibitem[{{Lerche}(1967)}]{lerche_67}
{Lerche}, I. 1967, \apj, 147, 689

\bibitem[{{Mao} \& {Ostriker}(2018)}]{mao_ostriker_18}
{Mao}, S.~A., \& {Ostriker}, E.~C. 2018, \apj, 854, 89

\bibitem[{{Mignone} {et~al.}(2018){Mignone}, {Bodo}, {Vaidya}, \&
  {Mattia}}]{mignone_18}
{Mignone}, A., {Bodo}, G., {Vaidya}, B., \& {Mattia}, G. 2018, \apj, 859, 13

\bibitem[{{Morlino} \& {Gabici}(2015)}]{morlino_gabici_15}
{Morlino}, G., \& {Gabici}, S. 2015, \mnras, 451, L100

\bibitem[{{Nava} {et~al.}(2016){Nava}, {Gabici}, {Marcowith}, {Morlino}, \&
  {Ptuskin}}]{nava_16}
{Nava}, L., {Gabici}, S., {Marcowith}, A., {Morlino}, G., \& {Ptuskin}, V.~S.
  2016, \mnras, 461, 3552

\bibitem[{{O'C Drury} {et~al.}(1996){O'C Drury}, {Duffy}, \& {Kirk}}]{drury_96}
{O'C Drury}, L., {Duffy}, P., \& {Kirk}, J.~G. 1996, \aap, 309, 1002

\bibitem[{{Padovani} {et~al.}(2020){Padovani}, {Ivlev}, {Galli}, {Offner},
  {Indriolo}, {Rodgers-Lee}, {Marcowith}, {Girichidis}, {Bykov}, \&
  {Kruijssen}}]{padovani_20}
{Padovani}, M., {Ivlev}, A.~V., {Galli}, D., {et~al.} 2020, \ssr, 216, 29

\bibitem[{{Pais} {et~al.}(2018){Pais}, {Pfrommer}, {Ehlert}, \&
  {Pakmor}}]{pais_18}
{Pais}, M., {Pfrommer}, C., {Ehlert}, K., \& {Pakmor}, R. 2018, \mnras, 478,
  5278

\bibitem[{{Pakmor} {et~al.}(2016){Pakmor}, {Pfrommer}, {Simpson}, {Kannan}, \&
  {Springel}}]{pakmor_16}
{Pakmor}, R., {Pfrommer}, C., {Simpson}, C.~M., {Kannan}, R., \& {Springel}, V.
  2016, \mnras, 462, 2603

\bibitem[{{Pfrommer} {et~al.}(2017){Pfrommer}, {Pakmor}, {Schaal}, {Simpson},
  \& {Springel}}]{pfrommer_17}
{Pfrommer}, C., {Pakmor}, R., {Schaal}, K., {Simpson}, C.~M., \& {Springel}, V.
  2017, \mnras, 465, 4500

\bibitem[{{Recchia} {et~al.}(2016){Recchia}, {Blasi}, \&
  {Morlino}}]{recchia_16}
{Recchia}, S., {Blasi}, P., \& {Morlino}, G. 2016, \mnras, 462, 4227

\bibitem[{{Reville} \& {Bell}(2012)}]{reville_bell_12}
{Reville}, B., \& {Bell}, A.~R. 2012, \mnras, 419, 2433

\bibitem[{{Reville} {et~al.}(2021){Reville}, {Giacinti}, \&
  {Scott}}]{reville_21}
{Reville}, B., {Giacinti}, G., \& {Scott}, R. 2021, \mnras, 502, 4137

\bibitem[{{Reville} {et~al.}(2007){Reville}, {Kirk}, {Duffy}, \&
  {O'Sullivan}}]{reville_07}
{Reville}, B., {Kirk}, J.~G., {Duffy}, P., \& {O'Sullivan}, S. 2007, \aap, 475,
  435

\bibitem[{{Ruszkowski} {et~al.}(2017){Ruszkowski}, {Yang}, \&
  {Zweibel}}]{ruszkowski_17}
{Ruszkowski}, M., {Yang}, H.-Y.~K., \& {Zweibel}, E. 2017, \apj, 834, 208

\bibitem[{{Schroer} {et~al.}(2020){Schroer}, {Pezzi}, {Caprioli}, {Haggerty},
  \& {Blasi}}]{schroer_20}
{Schroer}, B., {Pezzi}, O., {Caprioli}, D., {Haggerty}, C., \& {Blasi}, P.
  2020, arXiv e-prints, arXiv:2011.02238

\bibitem[{{Shalaby} {et~al.}(2021){Shalaby}, {Thomas}, \&
  {Pfrommer}}]{shalaby_20}
{Shalaby}, M., {Thomas}, T., \& {Pfrommer}, C. 2021, \apj, 908, 206

\bibitem[{{Silsbee} \& {Ivlev}(2019)}]{silsbee_19}
{Silsbee}, K., \& {Ivlev}, A.~V. 2019, \apj, 879, 14

\bibitem[{{Skilling}(1971)}]{skilling_71}
{Skilling}, J. 1971, \apj, 170, 265

\bibitem[{{Skilling}(1975)}]{skilling_75a}
---. 1975, \mnras, 172, 557

\bibitem[{{Spitzer}(1978)}]{spitzer_78}
{Spitzer}, L. 1978, {Physical processes in the interstellar medium},
  doi:10.1002/9783527617722

\bibitem[{{Squire} {et~al.}(2021){Squire}, {Hopkins}, {Quataert}, \&
  {Kempski}}]{squire_2020}
{Squire}, J., {Hopkins}, P.~F., {Quataert}, E., \& {Kempski}, P. 2021, \mnras,
  502, 2630

\bibitem[{{Tagger} {et~al.}(1995){Tagger}, {Falgarone}, \&
  {Shukurov}}]{tagger_95}
{Tagger}, M., {Falgarone}, E., \& {Shukurov}, A. 1995, \aap, 299, 940

\bibitem[{{Thomas} \& {Pfrommer}(2019)}]{thomas_pfrommer_18}
{Thomas}, T., \& {Pfrommer}, C. 2019, \mnras, 485, 2977

\bibitem[{{van Marle} {et~al.}(2018){van Marle}, {Casse}, \&
  {Marcowith}}]{vanMarle_18}
{van Marle}, A.~J., {Casse}, F., \& {Marcowith}, A. 2018, \mnras, 473, 3394

\bibitem[{{Wentzel}(1969)}]{1969ApJ...156..303W}
{Wentzel}, D.~G. 1969, \apj, 156, 303

\bibitem[{{Wentzel}(1974)}]{wentzel_74}
---. 1974, \araa, 12, 71

\bibitem[{{Wiener} {et~al.}(2013){Wiener}, {Oh}, \& {Guo}}]{wiener_13}
{Wiener}, J., {Oh}, S.~P., \& {Guo}, F. 2013, \mnras, 434, 2209

\bibitem[{{Wiener} {et~al.}(2017){Wiener}, {Pfrommer}, \& {Oh}}]{wiener_17b}
{Wiener}, J., {Pfrommer}, C., \& {Oh}, S.~P. 2017, \mnras, 467, 906

\bibitem[{{Xu} {et~al.}(2016){Xu}, {Yan}, \& {Lazarian}}]{xu_2016}
{Xu}, S., {Yan}, H., \& {Lazarian}, A. 2016, \apj, 826, 166

\bibitem[{{Yan} \& {Lazarian}(2011)}]{yan_lazarian_11}
{Yan}, H., \& {Lazarian}, A. 2011, \apj, 731, 35

\bibitem[{{Yang} {et~al.}(2012){Yang}, {Ruszkowski}, {Ricker}, {Zweibel}, \&
  {Lee}}]{yang_12}
{Yang}, H. Y.~K., {Ruszkowski}, M., {Ricker}, P.~M., {Zweibel}, E., \& {Lee},
  D. 2012, \apj, 761, 185

\bibitem[{{Zirakashvili} {et~al.}(1996){Zirakashvili}, {Breitschwerdt},
  {Ptuskin}, \& {Voelk}}]{zirakashvili_96}
{Zirakashvili}, V.~N., {Breitschwerdt}, D., {Ptuskin}, V.~S., \& {Voelk}, H.~J.
  1996, \aap, 311, 113

\bibitem[{{Zweibel}(2013)}]{2013PhPl...20e5501Z}
{Zweibel}, E.~G. 2013, Physics of Plasmas, 20, 055501

\bibitem[{{Zweibel}(2017)}]{zweibel_17}
---. 2017, Physics of Plasmas, 24, 055402

\bibitem[{{Zweibel} \& {Shull}(1982)}]{zweibel_shull_82}
{Zweibel}, E.~G., \& {Shull}, J.~M. 1982, \apj, 259, 859

\end{thebibliography}

\end{document}